# Effect of Dynamic Disorder on Charge Transport in Organic Molecules

K. Navamani and K. Senthilkumar
Department of Physics, Bharathiar University
Coimbatore-641 046, INDIA.

**Abstract**

The charge transfer integral, site energy and the stacking angle fluctuations are used to study the hole and electron transport in recently synthesized dialkyl substituted thienothiophene caped benzobisthiazole (BDHTT-BBT) and methyl-substituted dicyanovinyl-capped quinquethiophene (DCV5T-Me) molecules. The charge transfer parameters, such as coherent and incoherent rate coefficients, hopping conductivity, mobility, disorder drift time, drift force, potential equilibrium rate and density flux rate are calculated and discussed. It has been observed that the charge decay up to the crossover point (or disorder drift time) is exponential, non-dispersive and charge transport follows the band-like transport. Beyond the disorder drift time, the charge decay is not fully exponential, dispersive and it follows the incoherent hopping transport. The proposed expressions for density flux and diffusion shows their dependency on dynamic disorder and is in agreement with the Troisi's model on diffusion limited by thermal disorder. The density flux rate is directly related with the drift force which facilitates the charge transfer. Calculated electron hopping conductivity in the BDHTT-BBT and DCV5T-Me is 0.8 and 0.18 S/cm, respectively. Molecule BDHTT-BBT has good electron mobility of 0.36 $cm^2/V$ s, which has larger electron density flux rate and drift force of 1.7 $x10^{20}$ $C/m^3$ s and $1.44x10^{-12}$ N.

## 1. Introduction

Over the last few decades the organic electronics is an emerging field in science and technology due to its application in organic semiconducting devices, such as field-effect transistors,[1-3] light-emitting diodes,[4,5] solar cells[6,7] and in nanoscale molecular electronics.[8] The organic materials have soft condensed phase property and easily tunable electronic property through the structural modification and suitable functional group substitution.[9-13] In addition, they are environmentally friendly, mechanically flexible, having self-assembling character and the production cost is relatively low.[8,14,15] The weak intermolecular interaction, low dielectric permittivity and structural disorder in the organic materials increases the electron-phonon coupling which is responsible for localized electronic states.[8,16-19] Therefore, the thermally activated polaron hopping mechanism is used to describe the charge transfer (CT) process in the organic molecules[18,20,21] and the Marcus theory of charge transfer is used to study CT along the sequential localized sites.[14,22-24] Berlin et al.[25]studied the effect of static and dynamic fluctuations on charge transfer kinetics in the donor-bridge-acceptor systems and concluded that the dynamic fluctuation facilitates the non-dispersive band-like charge transport due to increase in self-averaging charge transfer integral. Böhlin et al.[26] found that the localized charge carrier on the dynamically disordered system does not significantly influenced by electron-phonon coupling. Troisi's charge transport models[19,20,23,27-29] for disordered systems show that the dynamical property of electronic and nuclear degrees of freedom leads to the intermediate charge transfer mechanism between localized hopping and delocalized band transport, and is termed as diffusion limited by dynamic disorder.

In our previous studies,[30,31] we found that the dynamic disorder facilitates the density flux along the sequential localized sites and is responsible for hopping conductivity. Kocherzhenko et al.[32,33] concluded that the charge carrier dynamics in the short range molecular order follows the coherent band model and in the long range the charge carrier follows incoherent hopping mechanism. Many experimental studies also evidentially support that the charge transport in the dynamically disordered system does not follow fully hopping or band-like.[32] That is, the charge transport in organic molecules follows the partially coherent band-like and partially incoherent hopping mechanism. The vibrational spectroscopic studies also shows that the band-like transport is possible in highly ordered organic crystals, like pentacene.[19] Here, the dynamic disorder by nuclear and electronic degrees of freedom dissipates the thermal energy and controls the charge diffusion process which makes transition from hopping to band-like transport. Therefore, this intermediate CT mechanism and the effect of nuclear and electronic degrees of freedom on CT is to be studied further to understand the CT in organic molecules. In the present study, we have proposed a CT model for localized hopping and delocalized band-like transport.

In the present study, we have studied the dynamic disorder effect on charge transport properties of recently synthesized dialkyl substituted thienothiophene caped benzobisthiazole (BDHTT-BBT) and methyl-substituted dicyanovinyl-capped quinquethiophene (DCV5T-Me) molecules.[34,35] The BDHTT-BBT molecule has electron deficient benzobisthiazole core attached with dialkyl substituted thienothiophene at the end positions. The presence of benzobisthiazole core provides the planarity, rigidity and strong π-π interaction. The experimental study[34] shows that in the crystal structure the molecules are packed as slipped parallel structure and the π-stacking distance is 3.52 Å. It has been observed that the dihedral angle between central benzobisthiazole core and dithienothiophene is 3.36º. The observed FET hole mobility is 0.003 $cm^2/Vs$.[34] The DCV5T-Me molecule has good intermolecular interactions with $\pi$-stacking distance of 3.28 Å.[35] The earlier experimental study shows that the methyl substitution at central thiophene ring improves the power conversion efficiency and provides the crystalline nature. The crystalline nature of DCV5T-Me provides better pathways for charge transport and in photo-voltaic *J-V* characteristic study measured current density is 11.5 $mA/cm^2$.[35]

In the present work, to study the charge transport along the $\pi$-stacked organic molecules, we proposed an expression for rate of density flux along the CT path, diffusion coefficient and time dependent potential energy. Also, the CT mechanism is studied on the basis of rate coefficient for coherent and incoherent CT, hopping conductivity, mobility, dispersive parameter and disorder drift time. A detailed theoretical formulation for expressions corresponding to density flux rate, charge carrier momentum distribution, time dependent potential difference and dynamic disorder dependent diffusion coefficient are given in next section and in Supporting information. The results obtained from theoretical calculations are discussed in results and discussion Section.

## 2. Theoretical formalism

By using tight binding Hamiltonian approach, the presence of excess charge in a $\pi$-stacked molecular system is expressed as[8,36]

$$\hat{H} = \sum_i \varepsilon_i(\theta) a_i^+ a_i + \sum_{i \neq j} J_{i,j}(\theta) a_i^+ a_j \tag{1}$$

where, $a_i^+$ and $a_i$ are the creation and annihilation operators, $\varepsilon_i(\theta)$ is the site energy, energy of the charge when it is localized at $i^{th}$ molecular site and is calculated as diagonal matrix element of the Kohn-Sham Hamiltonian, $\varepsilon_i = \langle \varphi_i | \hat{H}_{KS} | \varphi_i \rangle$. The second term of Equation (1), $J_{i,j}$ is the off-diagonal matrix element of the Hamiltonian, $J_{i,j} = \langle \varphi_i | \hat{H}_{KS} | \varphi_j \rangle$ known as charge transfer integral or electronic coupling which measures the strength of the overlap between $\varphi_i$ and $\varphi_j$ (HOMO or LUMO of nearby molecules *i* and *j*). Based on the semi-classical Marcus theory, the charge transfer rate $(k)$ is defined as,[22-24,30]

$$k = \frac{2\pi}{\hbar} \left| J_{eff} \right|^2 \rho_{FCT} \tag{2}$$

The effective charge transfer integral, $J_{eff}$ is defined in terms of charge transfer integral, $J$ spatial overlap integral, $S$ and site energy, $\varepsilon$ as,[37,38]

$$J_{eff\,i,j} = J_{i,j} - S_{i,j} \left( \frac{\varepsilon_i + \varepsilon_j}{2} \right) \tag{3}$$

where, $\varepsilon_i$ and $\varepsilon_j$ are the energy of a charge when it is localized at $i^{th}$ and $j^{th}$ molecules, respectively. The site energy, charge transfer integral and spatial overlap integral were computed using the fragment molecular orbital (FMO) approach as implemented in the Amsterdam Density Functional (ADF) theory program.[38-40] In ADF calculation, we have used the Becke-Perdew (BP)[41,42] exchange correlation functional with triple-ζ plus double polarization (TZ2P) basis set.[43] In this procedure, the charge transfer integral and site energy corresponding to hole and electron transport are calculated directly from the Kohn-Sham Hamiltonian.[8,38]

In Equation (2), the Franck-Condon (FC) factor, $\rho_{FCT}$ measures the weightage of density of states (DOS)[22,23,25] and is calculated from the reorganization energy (λ) and the site energy difference between final and initial states, $\Delta\varepsilon_{ij} = \varepsilon_j - \varepsilon_i$.

$$\rho_{FCT} = \frac{1}{\sqrt{4\pi\lambda k_B T}} \exp\left[ -\frac{(\Delta\varepsilon_{ij} + \lambda)^2}{4\lambda k_B T} \right] \quad (4)$$

The reorganization energy measures the change in energy of the molecule due to the presence of excess charge and changes in the surrounding medium. The reorganization energy due to the presence of excess hole (positive charge, $\lambda_+$) and electron (negative charge, $\lambda_-$) is calculated as,[8,44,45]

$$\lambda_{\pm} = \left[E^{\pm}(g^0) - E^{\pm}(g^{\pm})\right] + \left[E^0(g^{\pm}) - E^0(g^0)\right] \quad (5)$$

where, $E^{\pm}(g^0)$ is the total energy of an ion in neutral geometry, $E^{\pm}(g^{\pm})$ is the energy of an ion in ionic geometry, $E^0(g^{\pm})$ is the energy of the neutral molecule in ionic geometry and $E^0(g^0)$ is the optimized ground state energy of the neutral molecule. The geometry of the studied BDHTT-BBT, DCV5T-Me molecules in neutral and ionic states are optimized using density functional theory method, B3LYP[46-50] with the 6-31G(d,p) basis set, as implemented in the GAUSSIAN 09 package.[50]

As reported in previous studies,[8,25,39,51] the structural fluctuations in the form of periodic fluctuation in π-stacking angle strongly influence the rate of charge transfer. In the present study, we have performed the kinetic Monte-Carlo (KMC) simulation to study the charge carrier dynamics in studied molecules. During the simulation, the charge is propagated on the basis of rate of charge transfer calculated from Equation (2). Along with reorganization energy the effective charge transfer integral and site energy difference calculated for 300 conformation and the stacking angle ranges from 0 to 180$^o$ are given as the input to the KMC simulation to analyze the CT kinetics, and the structural fluctuation time is fixed as 0.1 ps. In this model, we assume that the charge transport is along the sequence of π-stacked molecules and the charge carrier does not reach the end of molecular chain within the time scale of simulation. In each step of KMC simulation, the most probable hopping pathway is found out from the simulated trajectories based on the charge transfer rate at particular conformation.

In the case of normal Gaussian diffusion of the charge carrier in one dimension, the diffusion constant, $D$ is calculated from mean square displacement, $\left\langle X^2(t)\right\rangle$ which increases linearly with time, $t$

$$D = \lim_{t\to\infty} \frac{\left\langle X^2(t)\right\rangle}{2t} \tag{6}$$

The charge carrier mobility is calculated from diffusion coefficient, $D$ by using the Einstein relation,[49]

$$\mu = \left(\frac{e}{k_B T}\right) D \tag{7}$$

The dynamic fluctuation effect on CT kinetics is characterized by using the rate coefficient, $k(t)$ which is defined as[25,30,51]

$$k(t) = -\frac{d \ln P(t)}{dt} \tag{8}$$

where, $P(t)$ is the survival probability of charge at particular site. Based on this analysis, the type of fluctuation (slow or fast) and corresponding non-Condon (NC) effect (kinetic or static) on CT kinetics is studied. The time dependency character of rate coefficient is analyzed by using the power law[25,30,51]

$$k(t) = k^a t^{a-1}, \qquad 0 < a \le 1 \tag{9}$$

The timely varying rate coefficient $k(t)$ is calculated by using Equation (8), and the rate coefficient, $k$ is calculated from the survival probability curve. The dispersive parameter, $a$ is calculated by fitting the plotted curve of rate coefficient versus time on the Equation (9) (see Figure 4). As described in our previous studies[30,31,51] the dynamic disorder effect is studied by using survival probability through the entropy relation.

$$\sum_t S(t) = -k_B \sum_t P(t) \log P(t) \tag{10}$$

The previous studies[25,51,52] show that the dynamic disorder kinetically drifts the charge carrier along the charge transfer path. The variation of disorder drift ($S(t)/k_B$) during charge transfer is numerically calculated by using the Equation (10). The disorder drift time, $S_t$ is the time at which the disorder drift is maximum and is calculated from the graph (see Figure 5). The timely varying disorder drift curve will provide the information about charge diffusion process and its dependency on dynamic disorder.

The dynamic disorder dependent density flux equation is written as[30]

$$\rho_S = \rho_{S_0} \exp\left( -\frac{3S(t)}{5k_B} \right) \tag{11}$$

where, $\rho_{S_0}$ is the density flux in the absence of dynamic disorder.

The hopping conductivity is calculated on the basis of density flux model and is described as,[30,31,53]

$$\sigma_{Hop} = \frac{3}{5} \varepsilon \frac{\partial P}{\partial t} \tag{12}$$

where, $\varepsilon$ is electric permittivity of the medium and $\frac{\partial P}{\partial t}$ is the rate of transition probability (or charge transfer rate). In agreement with the previous Hall-effect measurement studies by Podzorov et al.,[19,54] Equation (12) shows that the hopping conductivity depends only on the electric component of the medium, not depends on magnetic component (see Reference 30).

To get further insight on charge carrier dynamics in the dynamically disordered system, we have formulated the expressions for density flux rate, time dependent momentum distribution factor, time dependent potential difference and dynamic disorder dependent diffusion coefficient (see Equations 13-17). The above parameters will provide the information about charge distribution speed along the CT path, potential equilibrium rate due to drift force and charge diffusion limiting behavior by the thermal disorder.

The earlier studies[31,32] show that the dynamic disorder perturbs the localized charge carrier which is responsible for density flux. Here, the rate of change of density flux along the CT path is described on the basis of perturbed charge carrier density, $\rho$ and the drift force, $F_D$ as (see Equations S1-S3),

$$\frac{\partial \rho}{\partial t} = \frac{3}{\hbar(3\pi^2 e)^{1/3}} \rho^{2/3} F_D \qquad (13)$$

where, $e$ is the electronic charge, the drift force, $F_D$ is responsible for charge carrier diffusion by the dynamic disorder and is equal to the rate of change of charge carrier momentum.

As described in previous studies,[30,31,51] the dynamic disorder is drifting the charge carrier from one localized site to the next site and is analyzed by the disorder drift time ($S_t$). In the present study, the variation of momentum distribution of the charge carrier along the CT path with respect to time is calculated by using the following equation (see Equations S3-S12),

$$P(t)_{mom} = P_{mom,0}\left(\exp(1 - P(t))\right)^{1/5} \tag{14}$$

where, $P_{mom,0}$ is the momentum distribution in the absence of dynamic disorder, and $P(t)$ is the survival probability of the charge carrier. The above Equation (14) provides the information on changes of the charge distribution speed (see Figure 7), during the KMC simulation.

In the present study, charge transfer process is the thermal diffusion because no external electric field is applied for CT. The presence of excess charge at one end of π-stacked molecular chain introduces the potential difference, $V_d$. During the CT, the charge diffusion will occur up to the point where the potential equilibrium is reached, that is, $V_d = 0$. The change in potential with respect to time is defined in terms of survival probability, $P(t)$ as (see Equations S13-S19),

$$V_d(t) = \frac{k_B T}{e}\left(1 - \left(\exp(1 - P(t))\right)^{2/5}\right) \tag{15}$$

where, $k_B T$ is the thermal energy which is responsible for thermal diffusion. The above Equation (15) gives the information about the potential equilibrium speed of the studied π-stacked molecule, by thermal energy (see Figure 7).

The time evolution of potential difference is calculated from the solution of Poisson equation and is expressed as (see Equation S20-S24),

$$\left|\frac{\partial V_d}{\partial t}\right| = \frac{\rho}{\varepsilon} D \tag{16}$$

where, $\rho$ is the contributed π-electron density for charge transport. From the mean squared displacement and time dependent potential distribution curves (see Figures 4 and 7), diffusion coefficient and potential equilibrium rate are calculated, which are used in Equation (16) to find the charge density $(\rho)$. The calculated charge density is used to calculate the momentum of the charge carrier through Equation S1, which is used in Equation (14) to compute the momentum distribution with respect to time (see Figure 7). From this momentum distribution curve, the rate of momentum distribution (drift force, $F_D$) is calculated. Here, the calculated drift force ($F_D$) and perturbed charge density $(\rho)$ are used in Equation (13) to study the density flux rate for hole and electron transport in the studied molecules.

The previous studies[30-32] show that the dynamic disorder makes the interaction with localized energy states, like as perturbation, which is responsible for density flux and existence of degeneracy levels. Here, the charge diffusion is controlled by the dynamic disorder. In this paper, the dynamic disorder dependent diffusion coefficient, $D_S$ is expressed as (see Equations S25-S36)

$$D_S = D_{S_0} \exp\left( -\frac{2S(t)}{5k_B} \right) \tag{17}$$

where, $D_{S_0}$ is the diffusion coefficient in the absence of dynamic disorder. The Equation (17) is in agreement with the Troisi's model on diffusion limited by dynamic disorder.[19,20,27,29,52]

To get the quantitative insight on charge transport in the presence of dynamic disorder, the information about stacking angle and its fluctuation around the equilibrium is required. As reported in previous study,[30,51] the equilibrium stacking angle and its fluctuation were calculated by using molecular dynamics (MD) simulation. The molecular dynamics simulation was performed for stacked dimers with fixed intermolecular distance of 3.52 Å for molecule BDHTT-BBT and 3.28 Å for molecule DCV5T-Me using NVT ensemble at temperature 298.15 K and pressure $10^{-5}$ Pa, using TINKER 4.2 molecular modeling package[55,56] with the standard molecular mechanics force field, MM3.[57,58] The simulations were performed up to 10 ns with time step of 1fs, and the atomic coordinates in trajectories were saved in the interval of 0.1 ps. The energy and occurrence of particular conformation were analyzed in all the saved 100000 frames to find the stacking angle and its fluctuation around the equilibrium value.[30,51]

## 3. Results and Discussion

The geometry of the molecules BDHTT-BBT and DCV5T-Me is optimized using DFT method at B3LYP/6-31G(d,p) level of theory and is shown in Figure. S1. The charge transfer integral, spatial overlap integral and site energy corresponding to hole and electron transport are calculated based on orbital coefficients and energies of HOMO and LUMO. The density plot of HOMO and LUMO of the studied molecules calculated at B3LYP/6-31G(d,p) level of theory is shown in Figures. S2 and S3, respectively. It has been observed that the HOMO and LUMO are π orbital and are delocalized on the entire core of the studied molecules and there is no density on the alkyl side chains of BDHTT-BBT. That is, in the π-stacked molecules, the overlap of core region of nearby molecules will favor both hole and electron transport along the columnar axis, and these molecules may have ambipolar character.

### 3.1. Effective Charge Transfer Integral

The previous studies[38,39,51] show that the effective charge transfer integral, $J_{eff}$ strongly depends on π-stacking distance and π-stacking angle. The experimental result shows that the π-stacking distance is 3.52 Å for molecule BDHTT-BBT and 3.28 Å for molecule DCV5T-Me.

The $J_{eff}$ for hole and electron transport in the BDHTT-BBT and DCV5T-Me dimer is calculated by using Equation (3). The variation of $J_{eff}$ with respect to stacking angle is shown in Figure 2. The MD results show that the equilibrium stacking angle for molecules BDHTT-BBT and DCV5T-Me is 16º and 28º, respectively, and the stacking angle fluctuation is up to 10 to 15º from the equilibrium stacking angle value (see Figure. S4). Around the equilibrium stacking angle, the $J_{eff}$ value is 0.02 eV for hole transport and is 0.07 eV for electron transport in the BDHTT-BBT and for DCV5T-Me molecule, $J_{eff}$ is 0.03 eV and 0.05 eV for hole and electron transport, respectively. The change in $J_{eff}$ due to the stacking angle fluctuation is included while calculating the CT kinetic parameters through kinetic Monte-Carlo simulation, and is used to study the dynamic fluctuation effect on charge transport mechanism.

### 3.2. Site Energy Difference

Site energy difference is one of the key parameters that determines the rate of CT and is equal to the difference in site energy ($\Delta\varepsilon_{ij} = \varepsilon_j - \varepsilon_i$) of nearby π-stacked molecules. The site energy difference arises due to the conformational disorder, electrostatic interactions and polarization effects. The previous studies[30,59,60] show that the site energy difference $(\Delta\varepsilon_{ij})$ provides significant impact on charge carrier dynamics and is acting as the driving force for CT when $\Delta\varepsilon_{ij}$ is negative, and is acting as a barrier for CT when $\Delta\varepsilon_{ij}$ is positive. The change in site energy difference with respect to the stacking angle for hole and electron transport in the studied molecules is shown in Figure 3. For hole and electron transport, within the equilibrium stacking angle fluctuation range, the BDHTT-BBT molecule has the $\Delta\varepsilon_{ij}$ of around 0.18 and -0.01 eV, respectively. At the equilibrium stacking angle, there is no site energy difference for hole transport and is 0.1 eV for electron transport in the molecule DCV5T-Me. Notably, no site energy difference is observed at the entire range of stacking angles for hole transport in molecule DCV5T-Me (see Figure 3). The calculated $\Delta\varepsilon_{ij}$ values at different stacking angles were included while calculating the CT rate and other kinetic parameters through Monte-Carlo simulation.

### 3.3. Reorganization Energy

The change in energy of the molecule due to structural reorganization by the presence of excess charge will act as a barrier for charge transport. The geometry of neutral, anionic and cationic states of the studied molecules were optimized at B3LYP/6-31G(d,p) level of theory and the reorganization energy is calculated by using Equation (5).

It has been observed that the molecules BDHTT-BBT and DCV5T-Me have the reorganization energy of 0.33 and 0.29 eV for the presence of excess positive charge. By analyzing the optimized geometry of neutral and cationic states of BDHTT-BBT and DCV5T-Me, we found that the presence of positive charge alters the torsional angle between the thiophene rings (S3-C3-C8-C9) in the DCV5T-Me by 3.5º, and the torsional angle between the benzobisthiazole core and hexylthiophene end units changes by 2º in the BDHTT-BBT. Comparatively BDHTT-BBT has rigid benzobisthiazole core with fused dialkyl thiothiophene which is responsible for the minimum electron reorganization energy of 0.19 eV, and the molecule DCV5T-Me has electron reorganization energy of 0.29 eV.

### 3.4. Charge Carrier Dynamics

The calculated charge transport key parameters such as effective charge transfer integral, site energy difference, reorganization energy and structural fluctuation in the form of stacking angle distribution are used to study the charge carrier dynamics through the kinetic Monte Carlo simulations. As shown in Figures 4 and S5 the mean squared displacement $\langle X^2(t) \rangle$ of the charge carrier calculated from Monte-Carlo simulation is linearly increasing with time. As described in Section 2, the diffusion coefficient, $D$ is obtained as half of the slope of the line, and based on Einstein relation (Equation 7) the charge carrier mobility is calculated from the $D$. The calculated survival probability and disorder drift are shown in Figure 5. As described in previous studies,[30,31,51] the disorder drift time, $S_t$ is calculated from the disorder drift curve (see Figure 5 and S5). In our previous study[30,31] we have concluded that the disorder drift time ($S_t$) is acting as the crossover point (COP) between adiabatic band and non-adiabatic hopping transport.

In this work, both the band-like and hopping charge transport mechanisms are studied through survival probability and disorder drift curve. Numerically computed rate, $k(t)$ values by MC simulation is used to analyze the time dependency character of rate coefficient by using equation (8) and (9). From the survival probability curve, the calculated rate coefficient up to the disorder drift time ($S_t$) is $k_1$ and beyond that point is $k_2$. These rate coefficients ($k_1$ and $k_2$) are used in Equation (9) to calculate the dispersive parameters ($a_1$ and $a_2$). The calculated rate coefficients ($k_1$ and $k_2$) and corresponding dispersive parameters ($a_1$ and $a_2$) are summarized in Table 1. Based on dispersive parameters ($a_1$ and $a_2$), the rate coefficients $k_1$ and $k_2$ are referred as the coherent and incoherent rate coefficients, respectively, which is conceptually supported by the previous studies.[25,30,31,51] Calculated coherent and incoherent rate coefficients ($k_1$ and $k_2$), average rate coefficient $(k)$, dispersive parameters ($a_1$ and $a_2$), hopping conductivity $(\sigma)$, mobility $(\mu)$ and disorder drift time $(S_t)$ are summarized in Table 1, for hole and electron transport in the molecules BDHTT-BBT and DCV5T-Me. It has been observed that the charge decay up to the COP (or disorder drift time) is exponential, non-dispersive and hence charge transport follows the band-like transport. Beyond COP, the charge decay is not exactly exponential, partially dispersive and the charge transport turns from band to incoherent hopping transport.[20,32] It has been found that the charge transfer rate coefficient up to the COP $(k_1)$ is time independent, beyond the COP the rate coefficient $(k_2)$ is time dependent which is analyzed through dispersive parameter calculated by using Equations (8) and (9). For instance, the calculated rate coefficients $k_1$ and $k_2$ for electron transport in molecule BDHTT-BBT are $1.87 \times 10^{13}$ and $1.12 \times 10^{13}$/s and their dispersive parameters $a_1$ and $a_2$ are 0.96 and 0.64 (see Figure 6 and Table 1), respectively. In this case, the dispersive parameter $a_1$ is nearly 1 which indicates that the rate coefficient $k_1$ is non-dispersive, time independent, but, in the latter case the rate coefficient $k_2$ is dispersive and time dependent. The average rate coefficient, $k$ is used in Equation (12) to calculate the hopping conductivity. For BDHTT-BBT molecule, the fluctuation in stacking angle around the equilibrium value of 14$^o$ is in the range of 0-28$^o$ and the effective electron transfer integral is in the range of 0.02 - 0.07 eV. The calculated hopping conductivity for electron is 0.8 S/cm. For DCV5T-Me molecule, the hoping conductivity is 0.18 S/cm and mobility is 0.07 cm$^2$/V s. The molecule DCV5T-Me has lesser hole and electron transporting ability due to the higher reorganization energy of 0.33 and 0.29 eV. In this case, the calculated disorder drift time is 1860 and 184 fs. That is, the positive charge carrier spend longer time on a localized HOMO of DCV5T-Me which is responsible for lesser hole mobility of 0.016 cm$^2$/V s.

To get further insight on charge transport in the studied BDHTT-BBT and DCV5T-Me molecules, charge carrier momentum distribution $(P_m)$ and potential $(V_{ij})$ variation at particular site due to charge diffusion are calculated from kinetic Monte Carlo simulation by using Equations (14) and (15), respectively and are shown in Figures 7 and S8. Here, the momentum distribution provides the information about the speed of charge distribution along the π-stacked units. The drift force $(F_D)$ and potential equilibrium rate $\left(\frac{\partial V}{\partial t}\right)$ are calculated from the charge carrier momentum distribution and potential rate curves (Figure 7), and are summarized in Table (1). The drift force $(F_D)$ is the driving force for charge diffusion, and is inversely proportional to the disorder drift time, $S_t$. The potential equilibrium rate $\left(\frac{\partial V}{\partial t}\right)$ provides the information about the time required for reaching the potential equilibrium during the charge transfer process. That is, the charge diffusion will occur until the potential equilibrium is reached. After this equilibrium point, the potential energy difference between the adjacent molecular units is zero, at which the CT will not take place. The calculated potential equilibrium rate $\left(\frac{\partial V}{\partial t}\right)$ is used in Equation (16) to calculate the π-electron density. As given in Equation (13), the density flux rate is calculated by using drift force $(F_D)$ and density $(\rho)$. The density flux rate expression gives the knowledge about the contributed charge density, due to perturbed effect of dynamic disorder, for CT mechanism within hopping time which is related to current density gradient. The potential equilibrium rate gives the information on how the ion injected molecules quickly reaches the potential equilibrium within hopping time, which is associated with the ionic diffusion property of the molecules. The potential equilibrium rate is calculated from the time evolutions of potential distribution curve (see Figure 7) and is used in Equation (16) to study the perturbed localized charge density. As observed from Table (1), the molecule BDHTT-BBT has the larger value of potential equilibrium rate ($3.85\text{x}10^{11}$ V/s), drift force (~$1.44\text{x}10^{-12}$ N) and density flux rate ($1.71\text{x}10^{20}$ $C/m^3s$) for electron transport. Hence, the electron mobility in this molecule is high (0.36 $cm^2/V$ s). Note that in this case the disorder drift time ($S_t$) is lower (42 fs). That is, due to the larger drift force the charge carrier does not spend longer time on the particular localized site. The potential equilibrium rate and drift force for hole transport in the studied DCV5T-Me molecule is comparatively minimum, the values are $1.44 \times 10^{10}$ V/s and $5.8 \times 10^{-14}$ N, and hence the hole transporting ability of this molecule is poor (see Table 1). The above results clearly show that the dynamic disorder is responsible for density flux along the charge transfer path which turns from hopping to band-like charge transport mechanism and is in agreement with the previous studies.[19,20,25,30]

The calculated density flux and diffusion coefficient are shown in Figures (8) and (9), respectively, for electron transport in the BDHTT-BBT molecule. As shown in Figures (8) and (9), and also S9 and S10, the charge density and diffusion coefficient are decreasing up to the disorder drift time, $S_t$ (COP), beyond that the charge density and diffusion coefficient are increasing. The dynamic disorder increases up to the disorder drift time, and then the dynamic disorder is decreased which is shown in Figure 5. That is, the charge density and diffusion coefficient are decreasing when the dynamic disorder is increasing; and the above parameters are increasing when the dynamic disorder is decreasing (see Figures 5, 8 and 9). That is, the charge flux and diffusion are controlled by dynamic disorder. Therefore, the dynamic disorder controls the localized hopping mechanism and turns the band-like CT mechanism, which is in agreement with the earlier studies of Troisi's model on diffusion limited by the dynamic disorder.[19,20,27,29,52]

## 4. Conclusion

The charge transport properties of BDHTT-BBT and DCV5T-Me molecules are studied by using electronic structure calculations. The charge transfer integral, site energy and the stacking angle fluctuation are used to calculate the charge transfer kinetic parameters, such as coherent and incoherent rate coefficients, hopping conductivity, mobility, disorder drift time, drift force, potential equilibrium rate and density flux rate and are used to study the hole and electron transport in the BDHTT-BBT and DCV5T-Me molecules. The disorder drift time is acting as the crossover point between hopping and band transport mechanisms. It has been found that the charge decay up to the crossover point (disorder drift time) is exponential, non-dispersive and charge transport follows the band-like transport. Beyond the disorder drift time the charge decay is not fully exponential and charge transport follows incoherent hopping transport. The proposed density flux and diffusion expressions shows that the localized charge transport in these molecules is limited by dynamic disorder and is in agreement with the Troisi's model on diffusion limited by thermal disorder. Molecule, BDHTT-BBT has good electron mobility of 0.36 $cm^2/V$ s, which has larger electron density flux rate and drift force of $1.7x10^{20}$ $C/m^3s$ and $1.44x10^{-12}$ N, respectively.

**Acknowledgement:** The authors thank the Department of Science and Technology (DST), India for awarding research project under Fast Track Scheme.

**Table 1** Reorganization energy (λ), coherent rate coefficient ($k_1$), incoherent rate coefficient ($k_2$), average rate coefficient ($k$), dispersive parameter for before and after the crossover point ($a_1$ and $a_2$), hopping conductivity $(\sigma)$, mobility $(\mu)$, disorder drift time ($S_t$), drift force ($F_D$), potential equilibrium rate $\left(\frac{\partial V}{\partial t}\right)$ and density flux rate $\left(\frac{\partial \rho}{\partial t}\right)$ for hole and transport in BDHTT-BTT and DCV5T-Me molecules.

| Molecules | λ (eV) | $k_1$ (x10$^{12}$/s) | $k_2$ (x10$^{12}$/s) | $k$ (x10$^{12}$/s) | $a_1$ | $a_2$ | $\sigma$ (S/cm) | $\mu$ (cm$^2$/V s) | $S_t$ (fs) | $F_D$ (x10$^{-13}$ N) | $\frac{\partial V}{\partial t}$ (x10$^{10}$ V/s) | $\frac{\partial \rho}{\partial t}$ (x10$^{19}$ C/m$^3$ s |
|---|---|---|---|---|---|---|---|---|---|---|---|---|
| **Hole** | | | | | | | | | | | | |
| BDHTT-BBT | 0.29 | 1.64 | 1.47 | 1.56 | 0.9 | 0.62 | 0.083 | 0.038 | 592 | 1.23 | 3.25 | 1.27 |
| DCV5T-Me | 0.33 | 0.784 | 0.780 | 0.782 | 0.92 | 0.81 | 0.041 | 0.016 | 1860 | 0.58 | 1.44 | 0.61 |
| **Electron** | | | | | | | | | | | | |
| BDHTT-BBT | 0.19 | 18.7 | 11.2 | 15 | 0.96 | 0.64 | 0.8 | 0.362 | 42 | 14.37 | 38.55 | 17.1 |
| DCV5T-Me | 0.29 | 4.7 | 2.2 | 3.45 | 0.93 | 0.67 | 0.18 | 0.072 | 184 | 4.37 | 10.38 | 6.33 |

**Figure 1:** The chemical structure of molecules (a) BDHTT-BBT and (b) DCV5T-Me

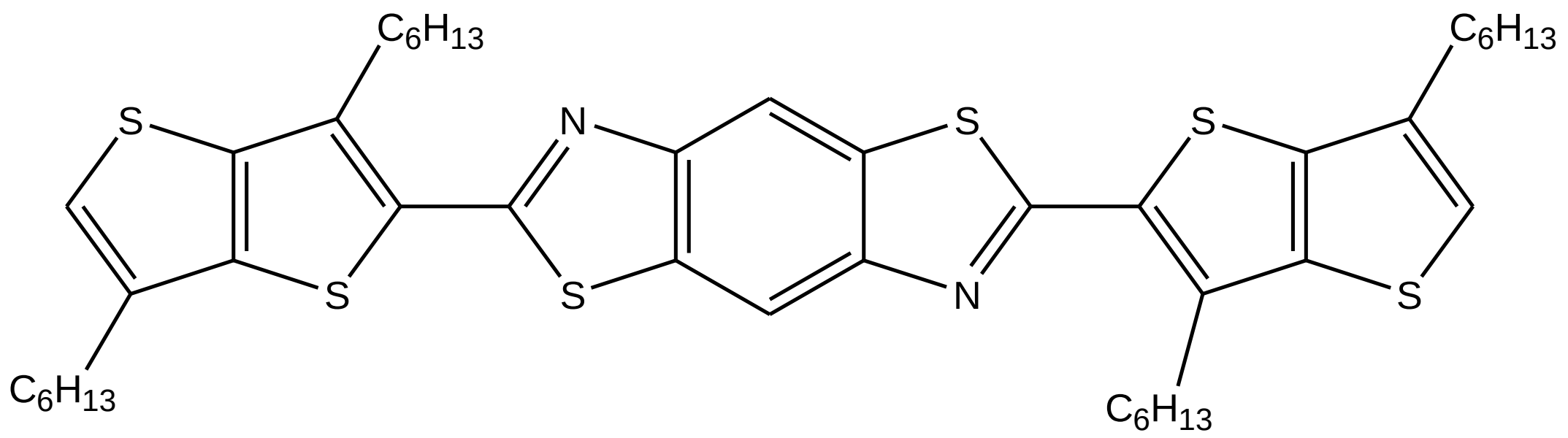

**(a)**

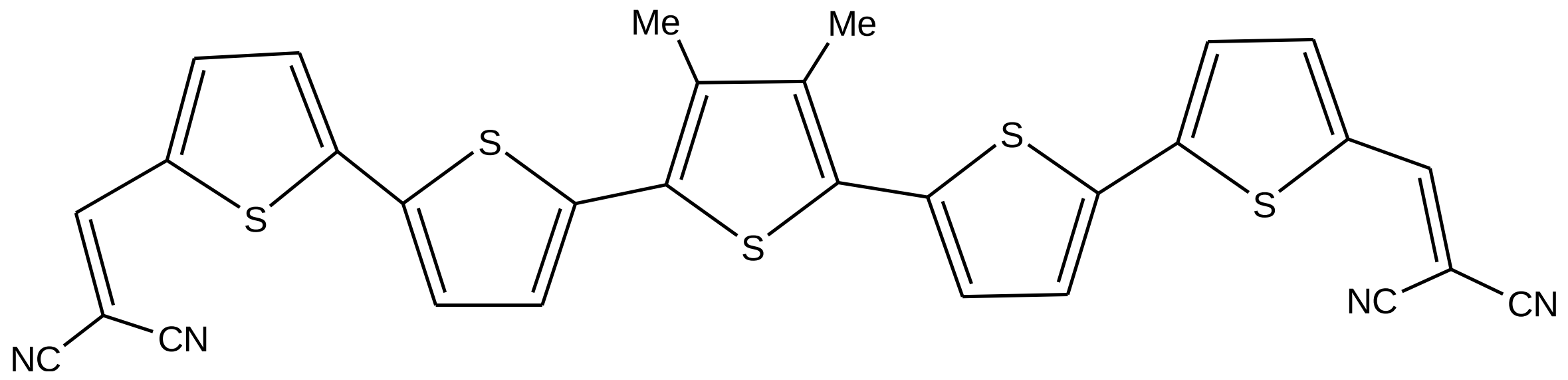

**(b)**

**Figure 2:** The effective charge transfer integral ($J_{eff}$, in eV) for hole (open circle) and electron (closed circle) transport in molecules BDHTT-BBT (dashed line) and DCV5T-Me (solid line) at different stacking angles ($\theta$, in degree)

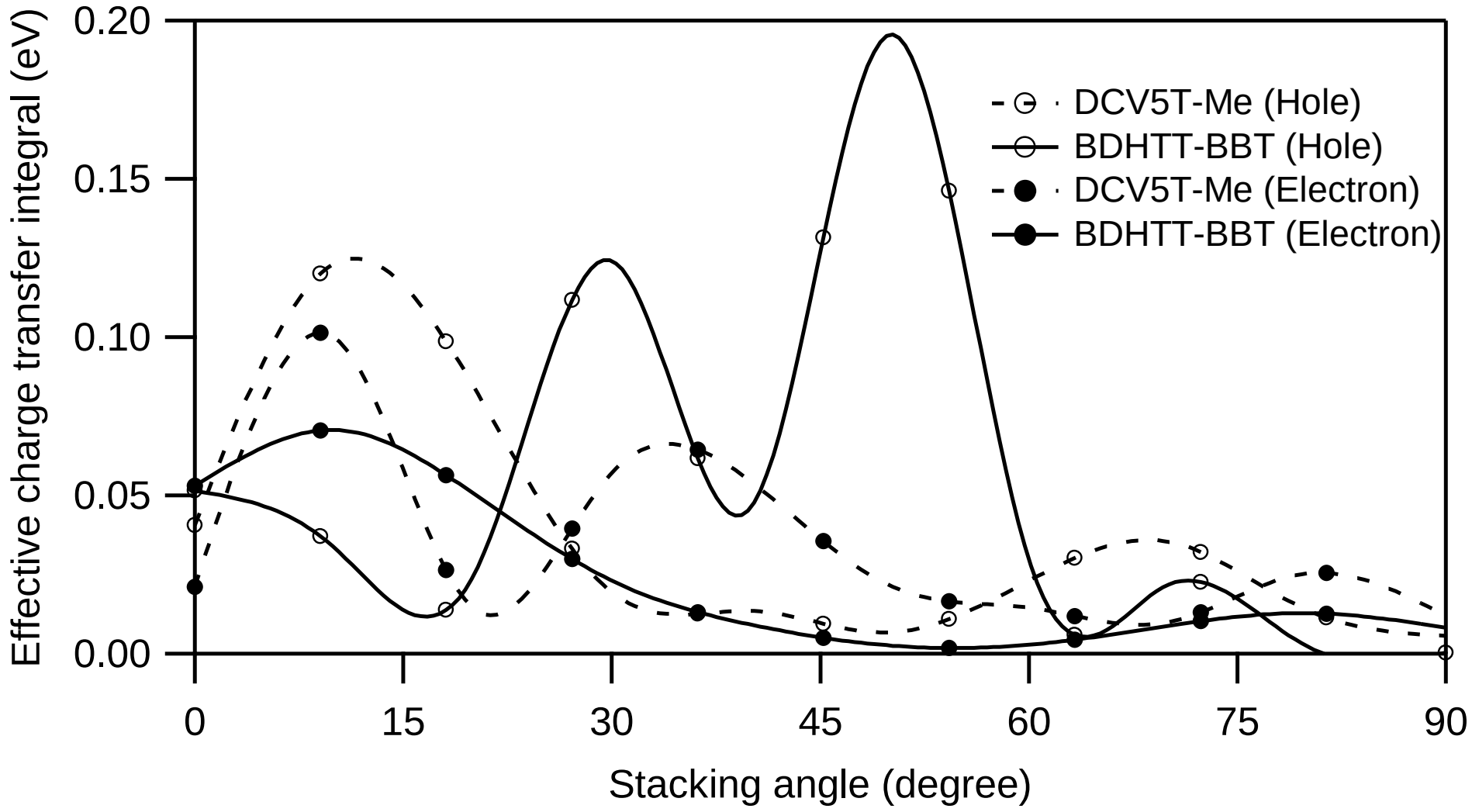

**Figure 3:** The site energy difference ($\Delta\varepsilon_{ij}$, in eV) for hole (open circle) and electron (closed circle) transport in molecules BDHTT-BBT (dashed line) and DCV5T-Me (solid line) at different stacking angles ($\theta$, in degree)

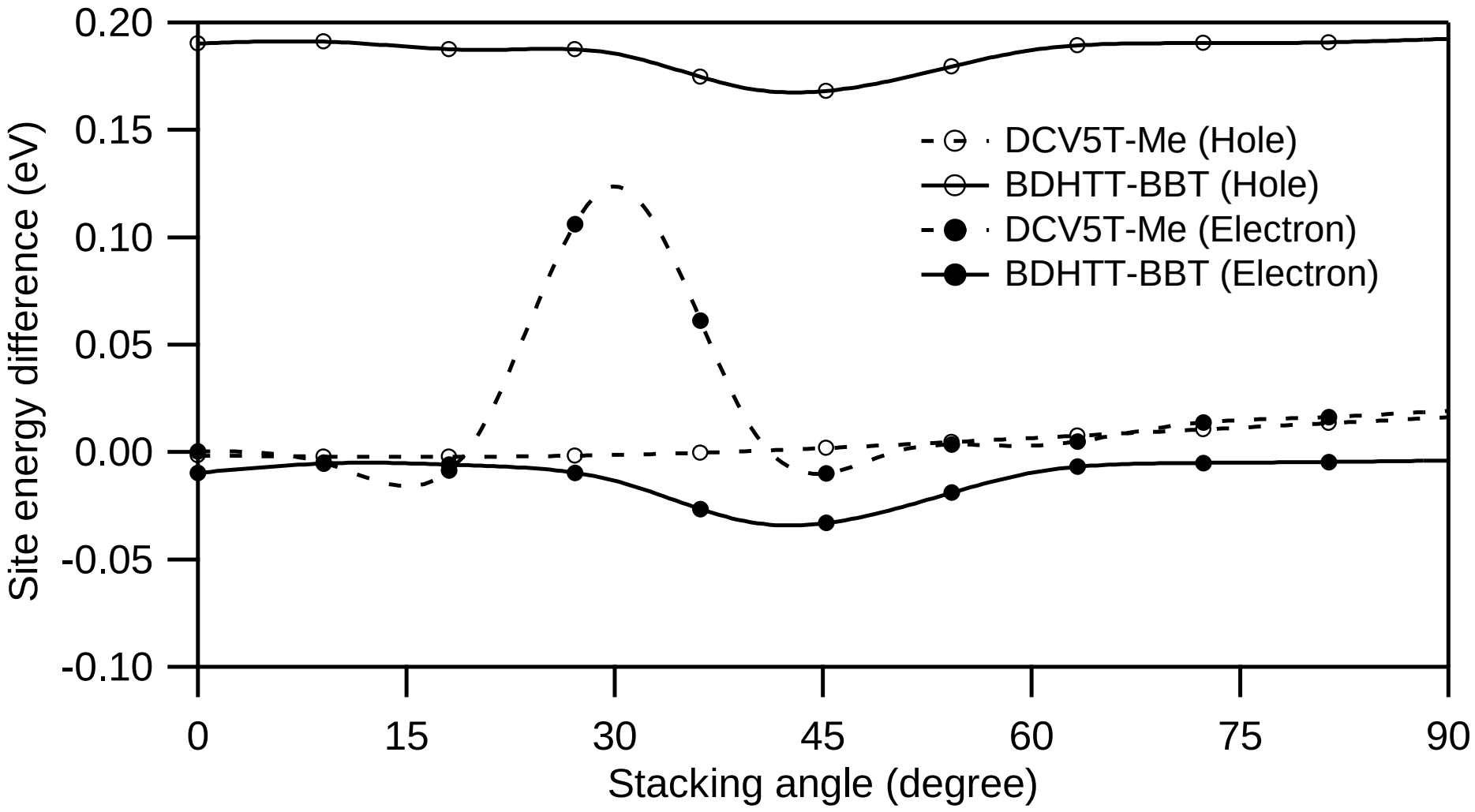

**Figure 4:** The mean squared displacement of negative charge in BDHTT-BBT molecule with respect to time.

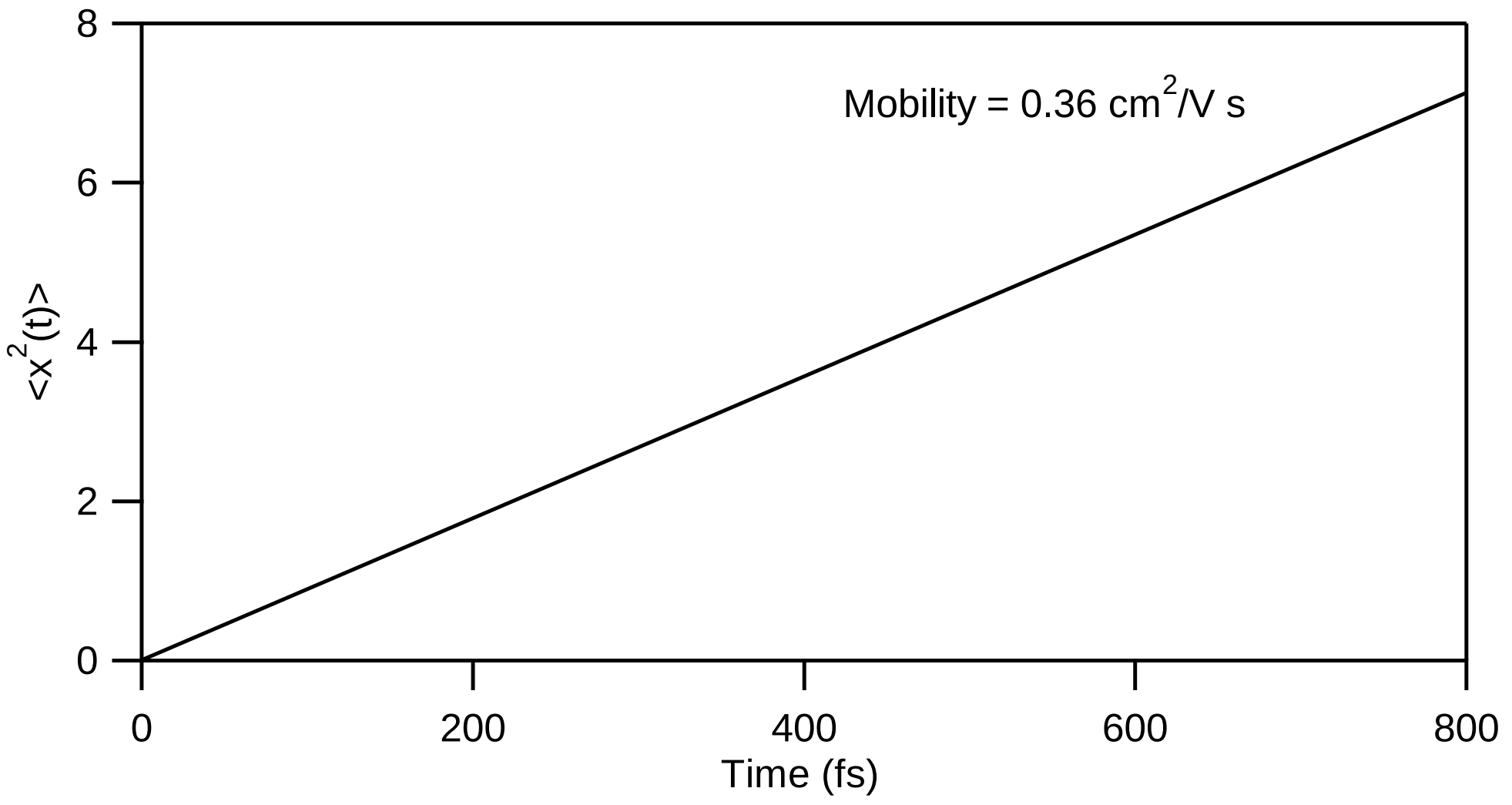

**Figure 5:** The survival probability and disorder drift of negative charge in BDHTT-BBT molecule with respect to time.

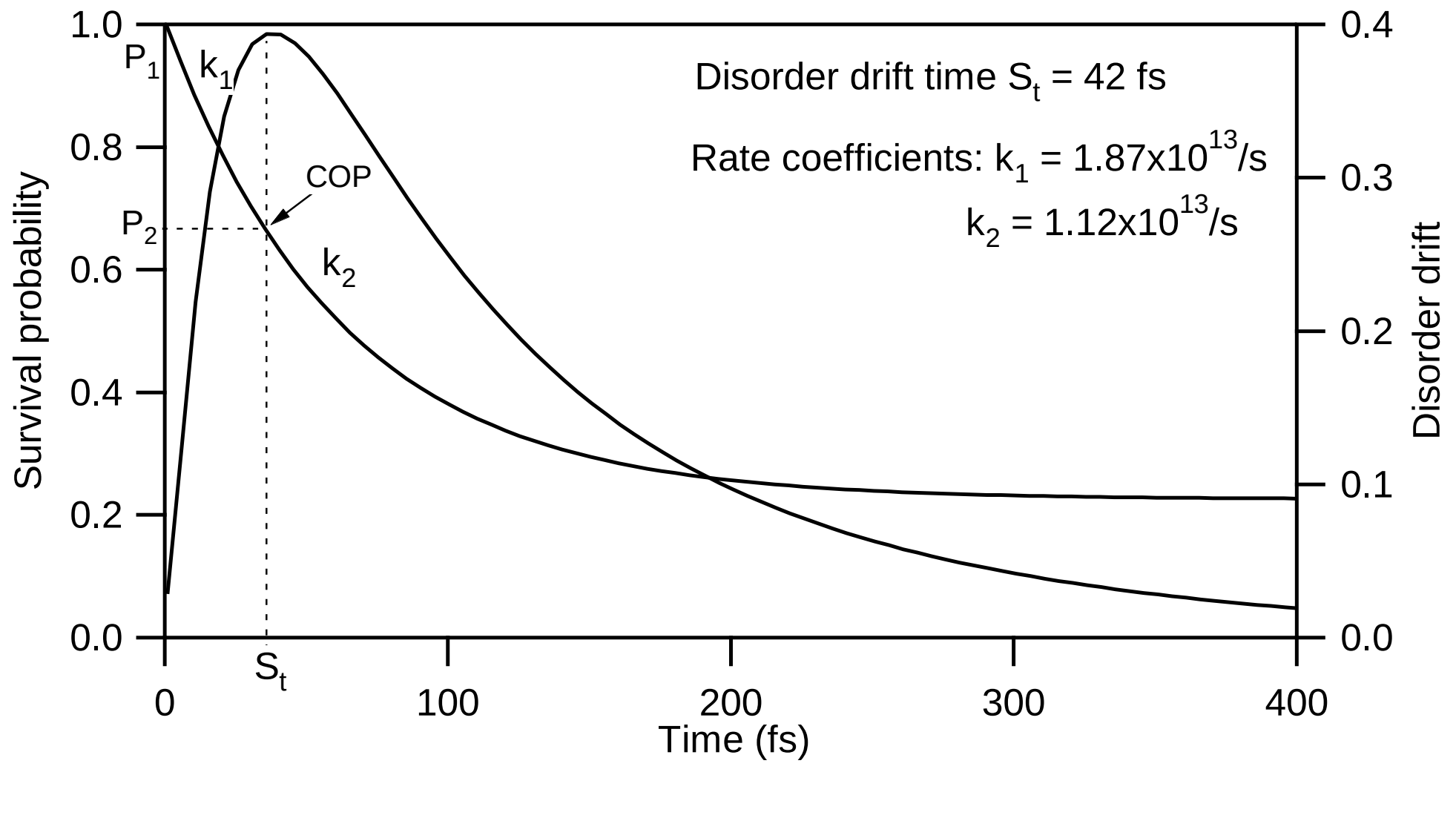

**Figure 6:** Time evolution of the rate coefficient for electron transport in BDHTT-BBT molecule in (a) coherent regime (b) non-coherent regime.

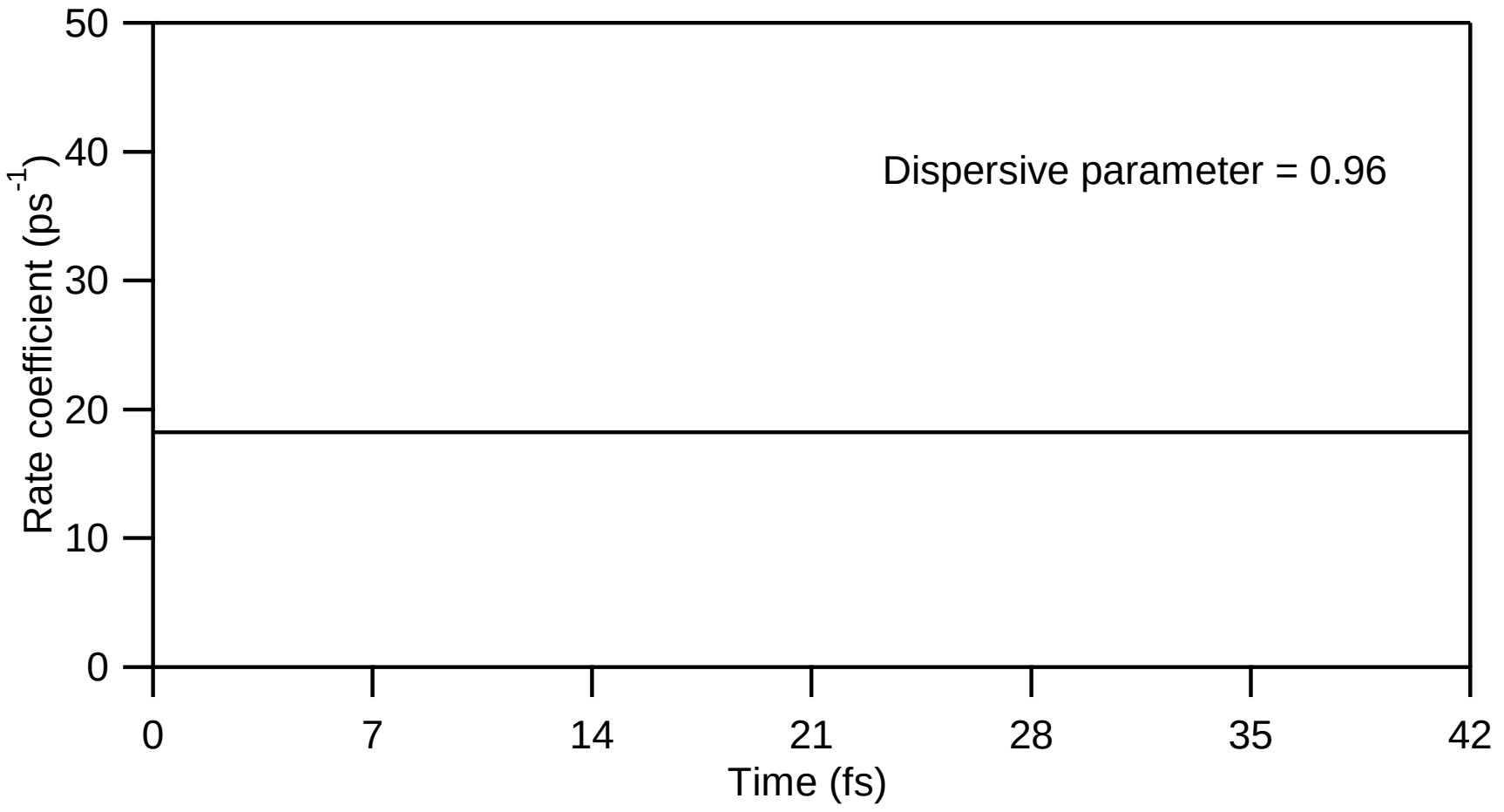

**(a)**

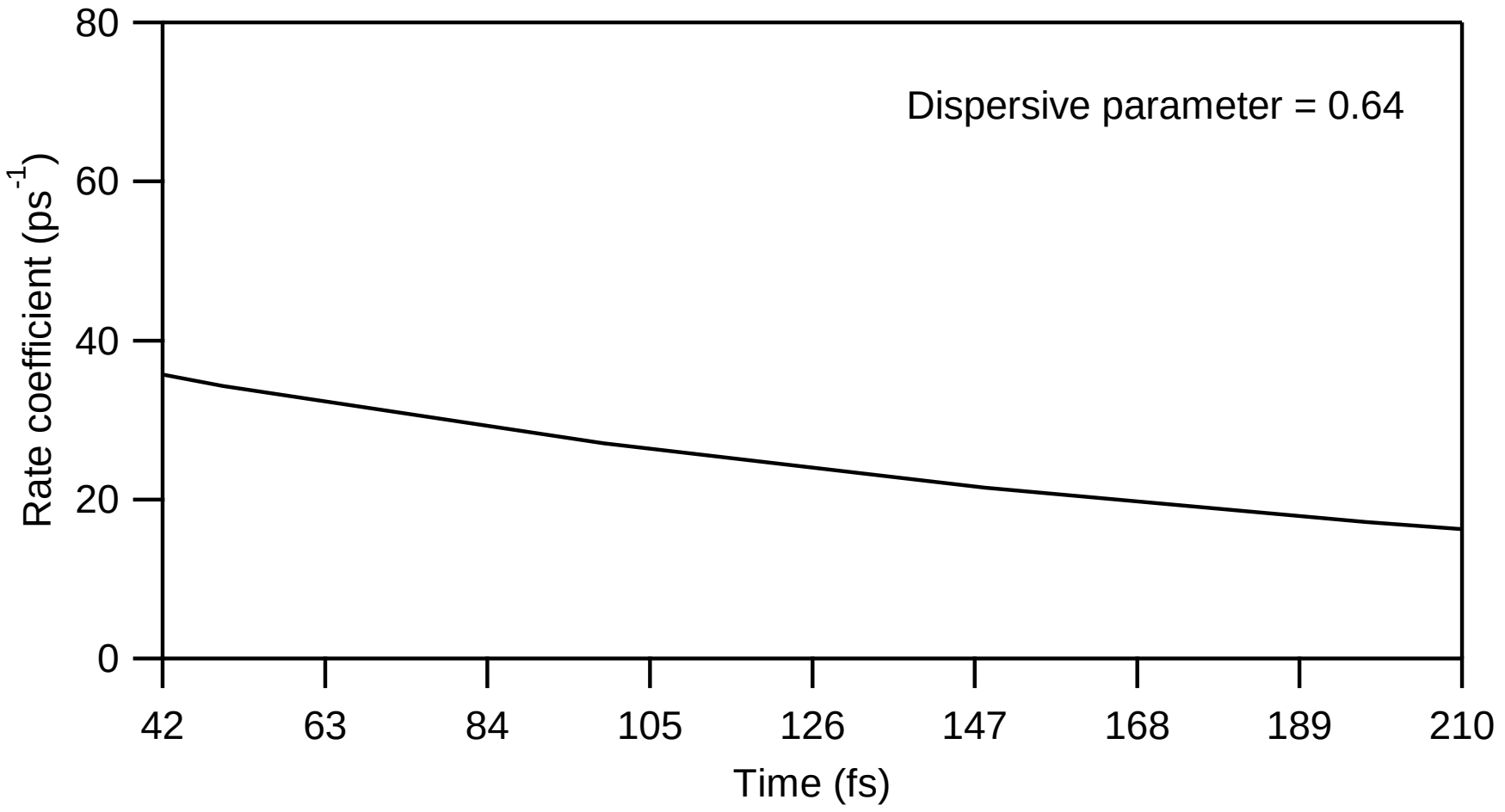

**(b)**

**Figure 7:** Time evolution of momentum ratio and potential distribution for electron transport in BDHTT-BBT molecule

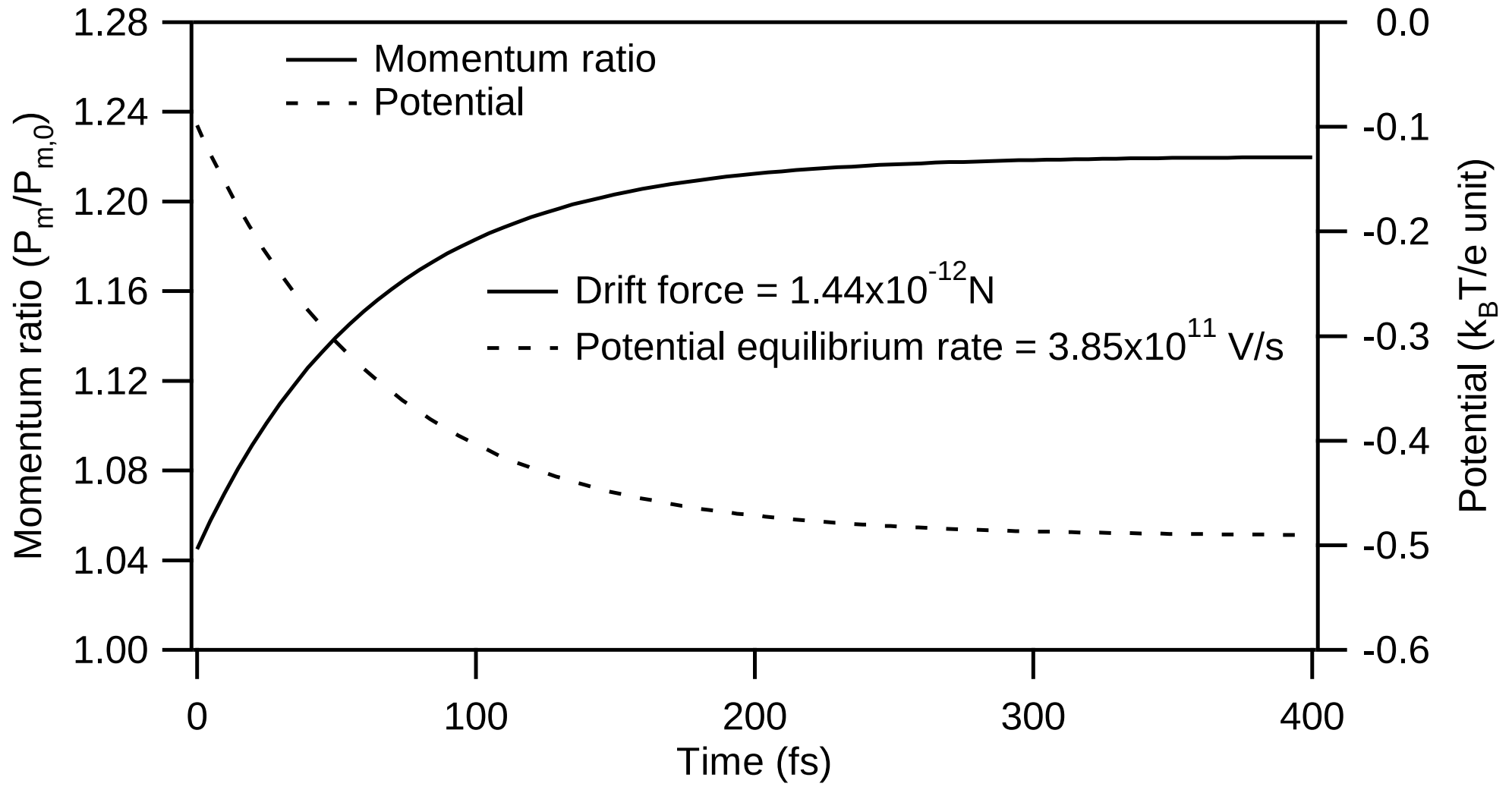

**Figure 8:** Time evolution of the density flux for electron transport in BDHTT-BBT molecule

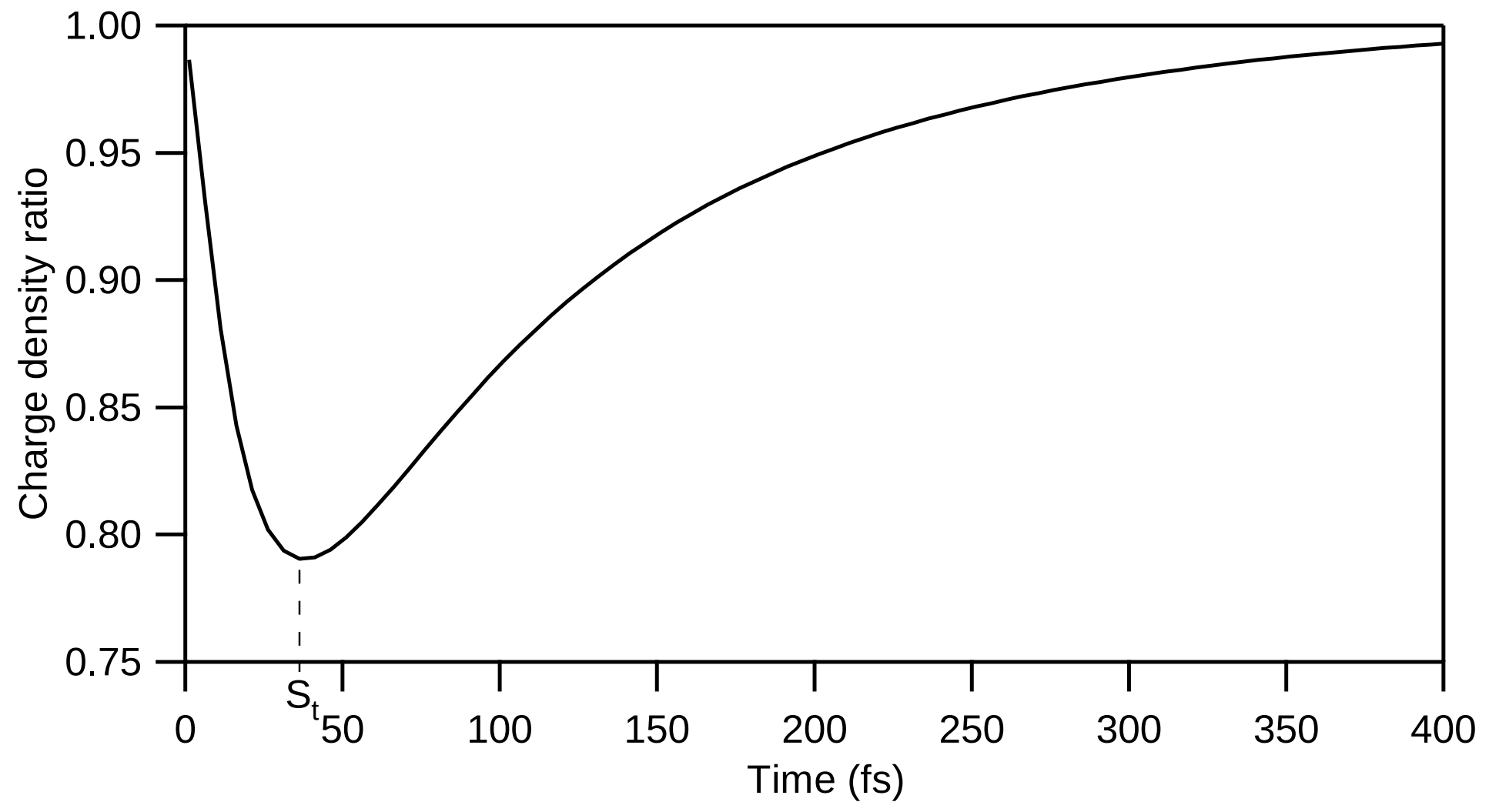

**Figure 9:** Time evolution of the diffusion coefficient for electron transport in BDHTT-BBT molecule

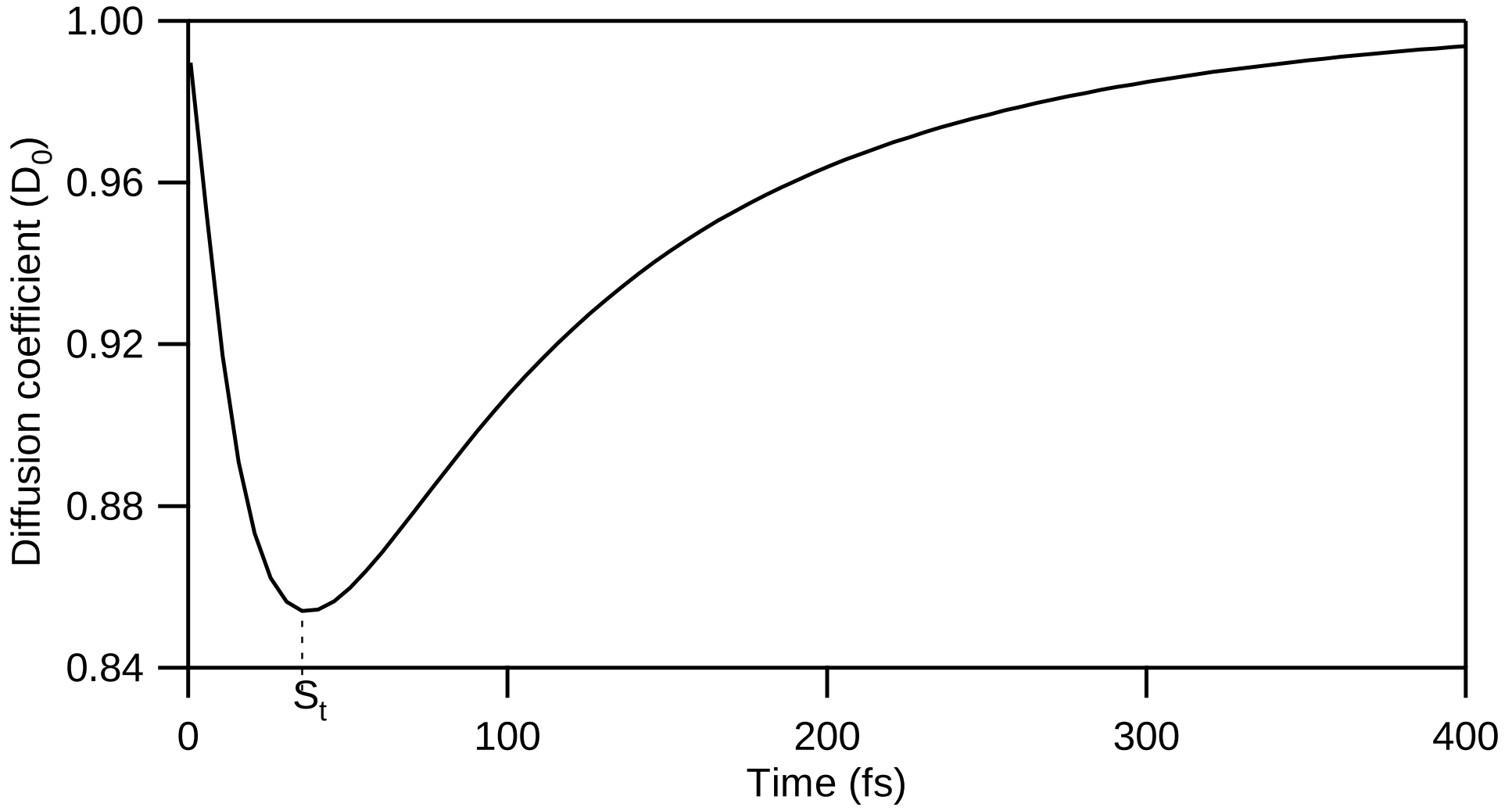

## Supporting Information

# Effect of Dynamic Disorder on Charge Transport in Organic Molecules

K. Navamani and K. Senthilkumar
Department of Physics, Bharathiar University
Coimbatore-641 046, INDIA.

**Figure S1.** Optimized structures of BDHTT-BBT, DCV5T-Me molecules

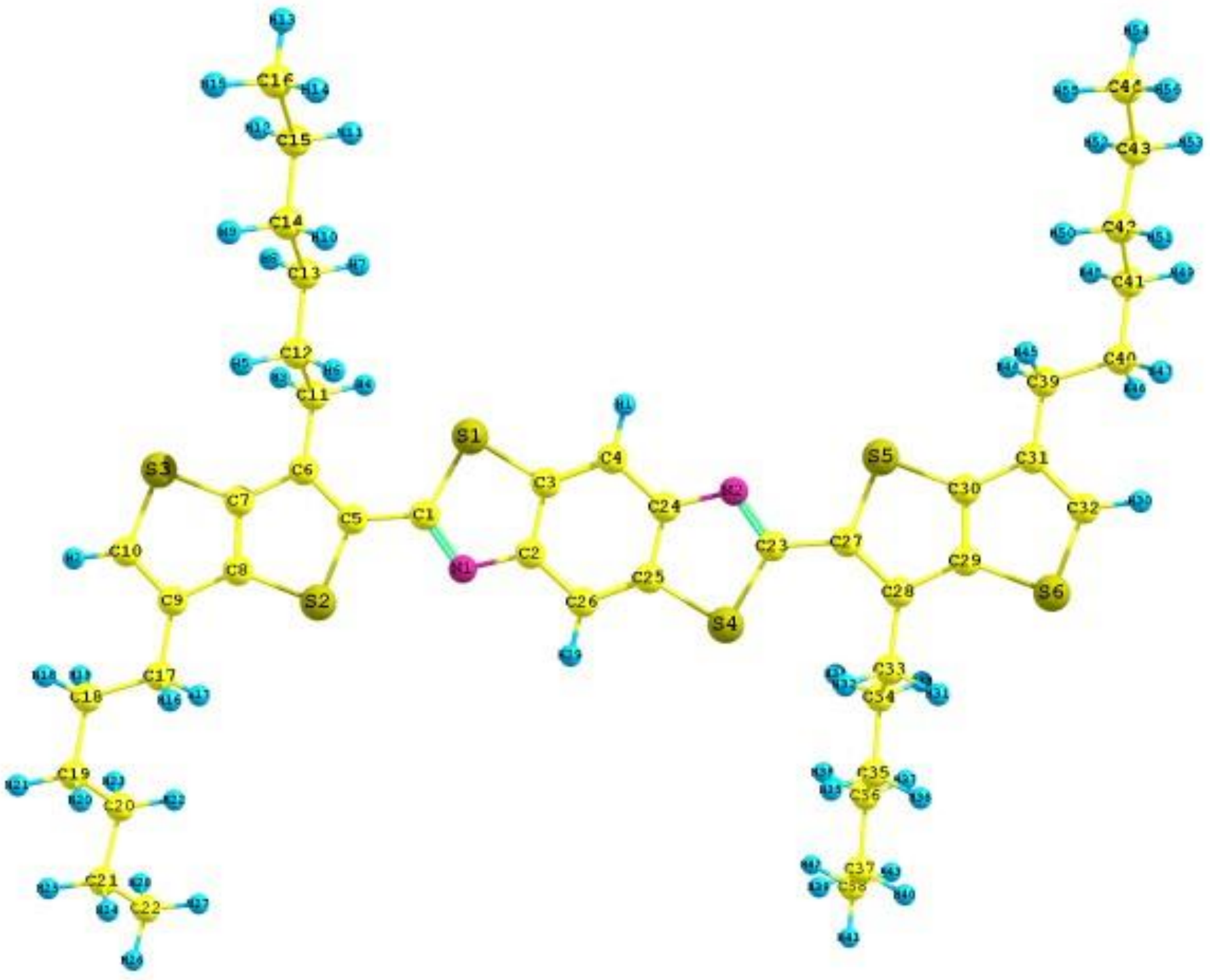

**BDHTT-BBT**

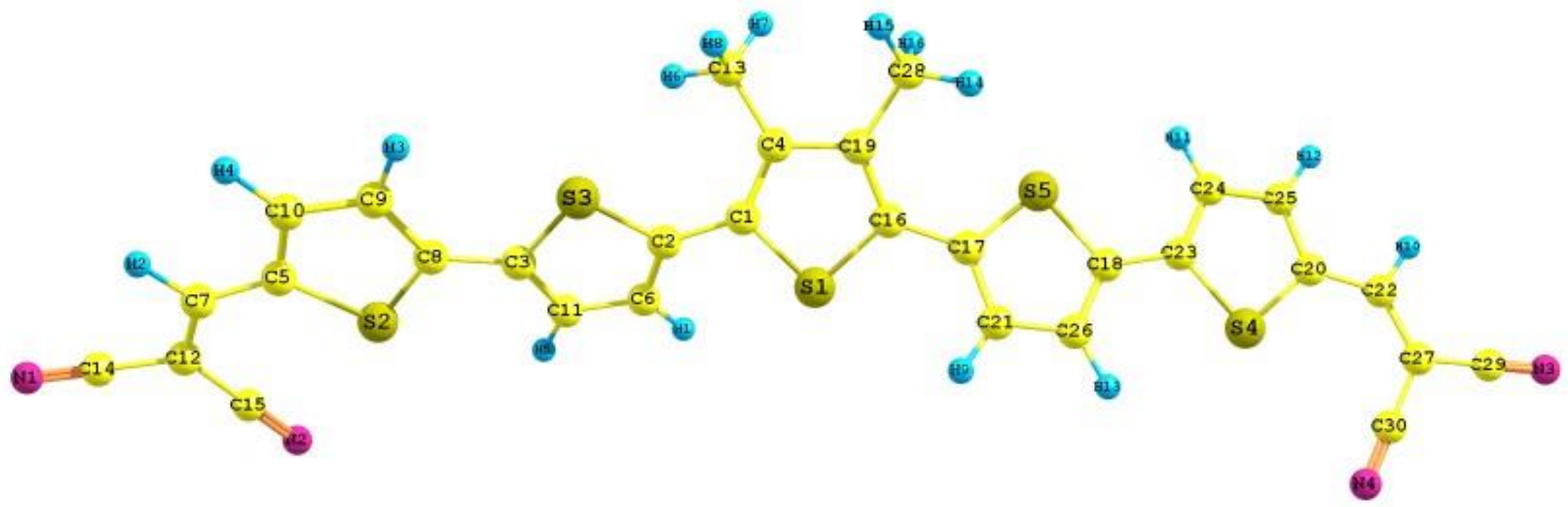

**DCV5T-Me**

**Figure S2.** The density plot of highest occupied molecular orbital (HOMO) of the studied BDHTT-BBT, DCV5T-Me molecules

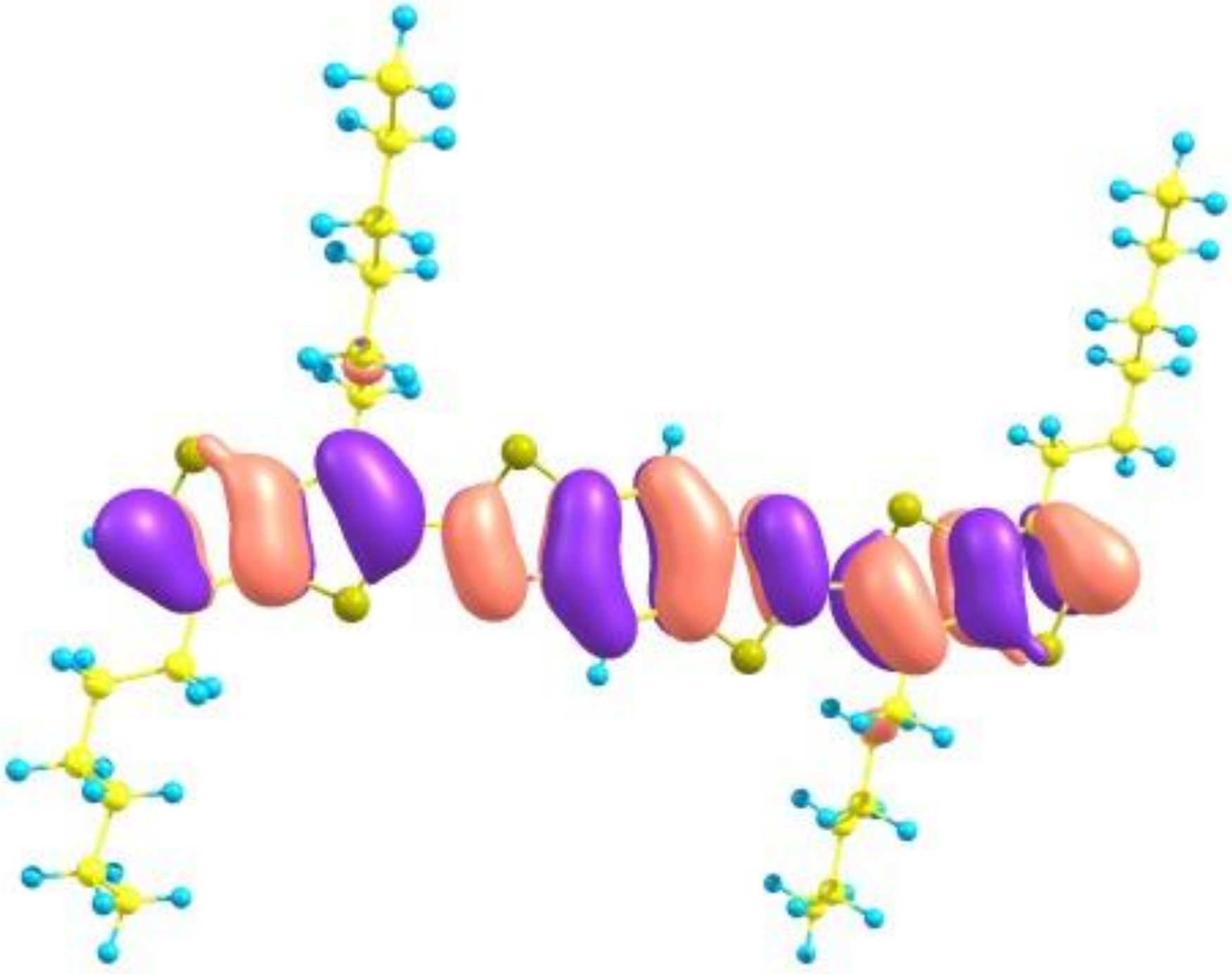

**BDHTT-BBT**

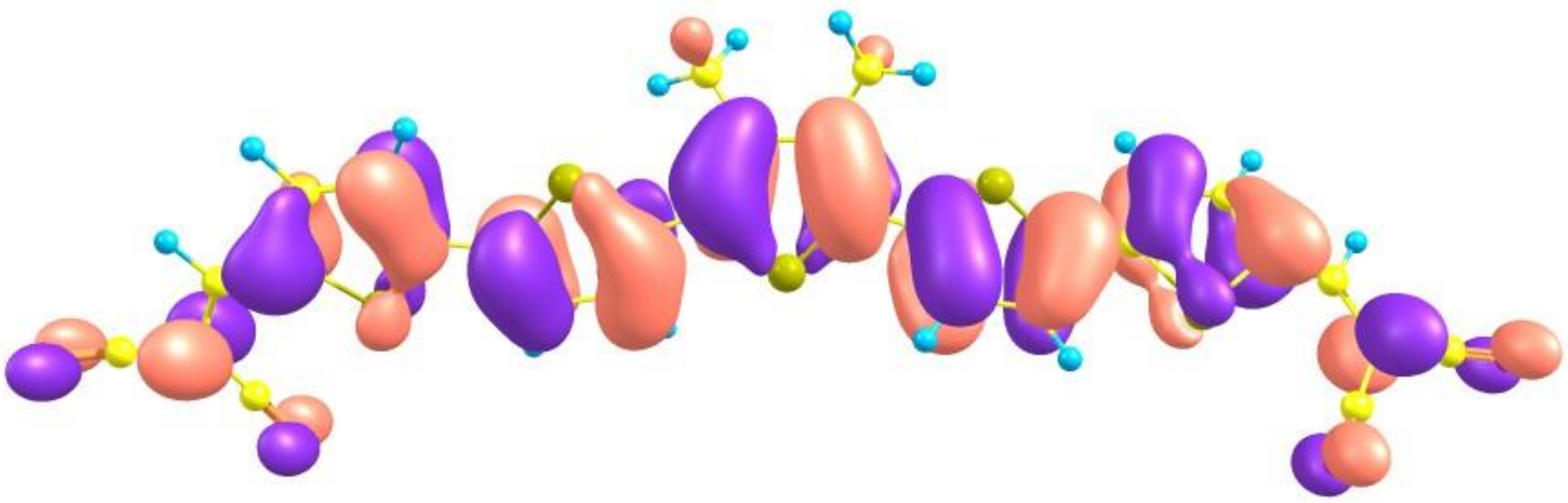

**DCV5T-Me**

**Figure S3.** The density plot of lowest unoccupied molecular orbital (LUMO) of the studied BDHTT-BBT, DCV5T-Me molecules

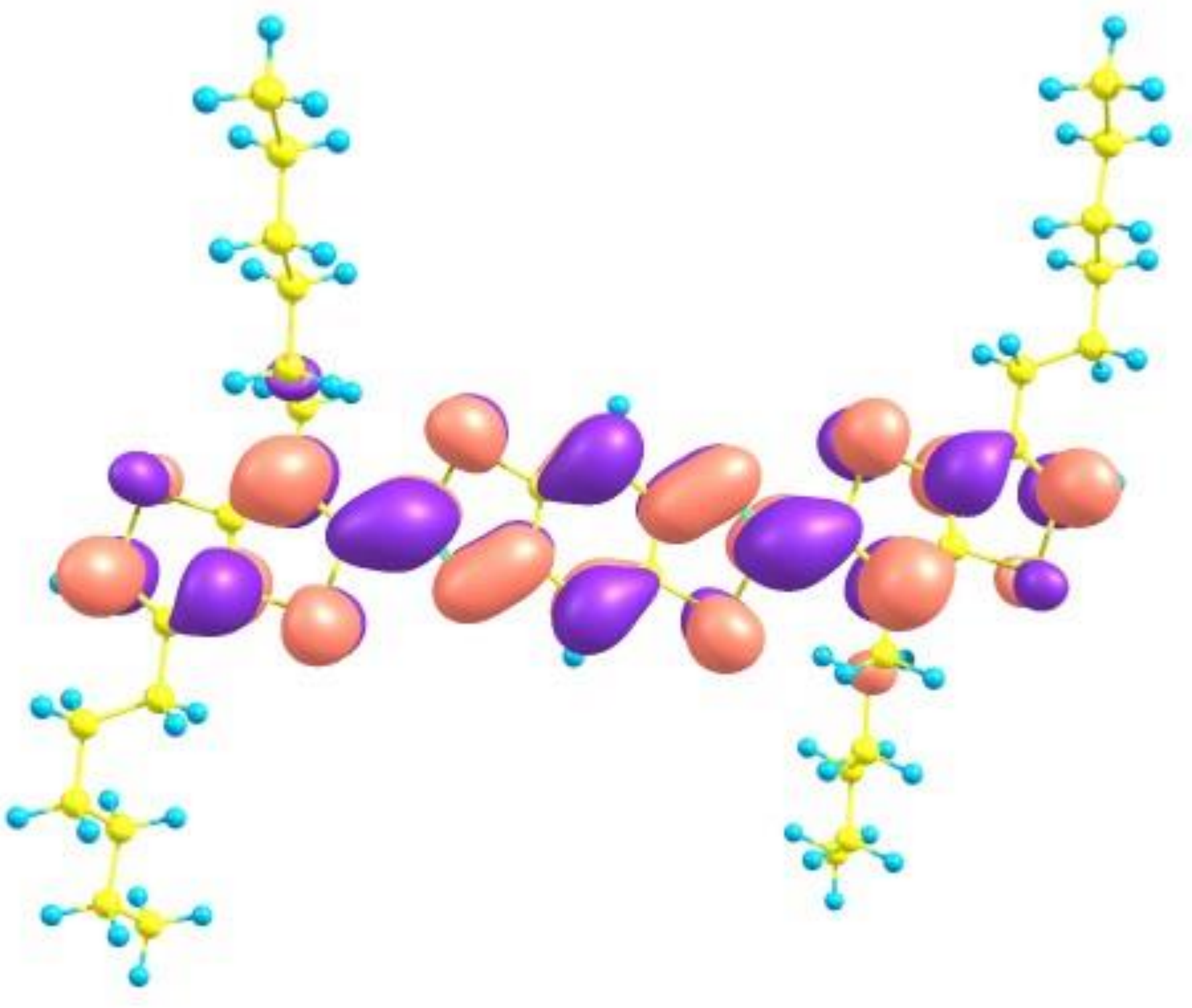

**BDHTT-BBT**

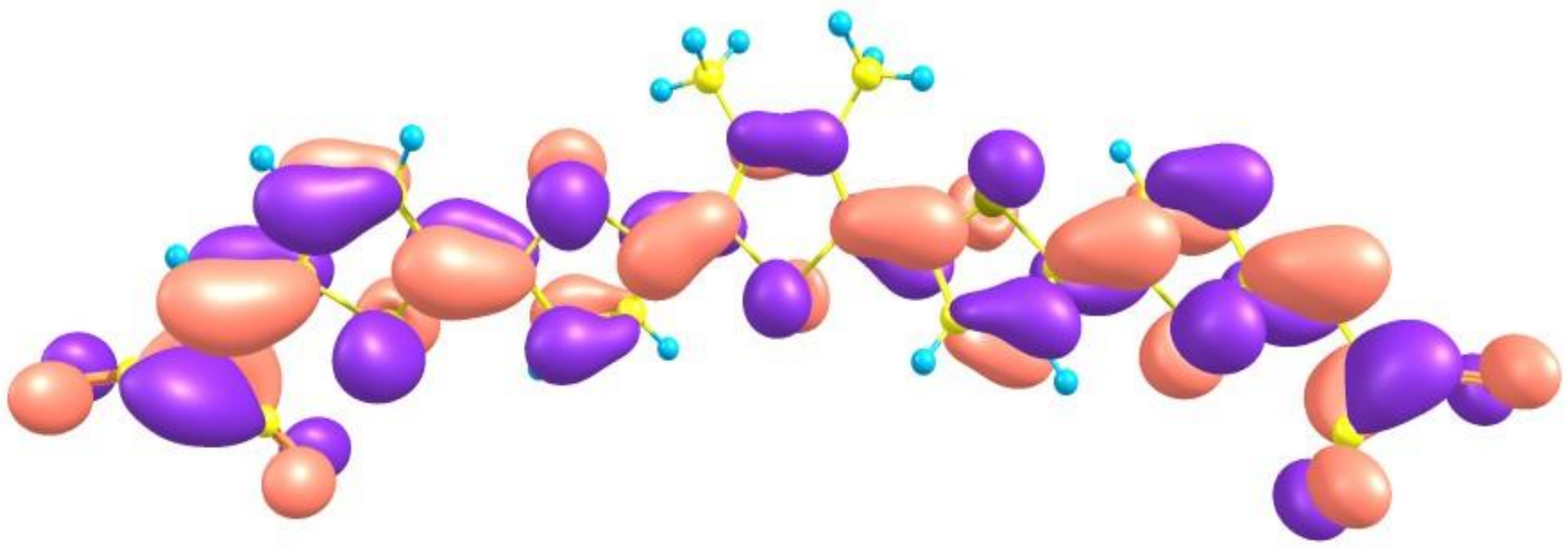

**DCV5T-Me**

**Figure S4. (a)** The plot between the number of occurrence, potential energy with respect to stacking angle calculation from molecular dynamics simulation for BDHTT-BBT molecule

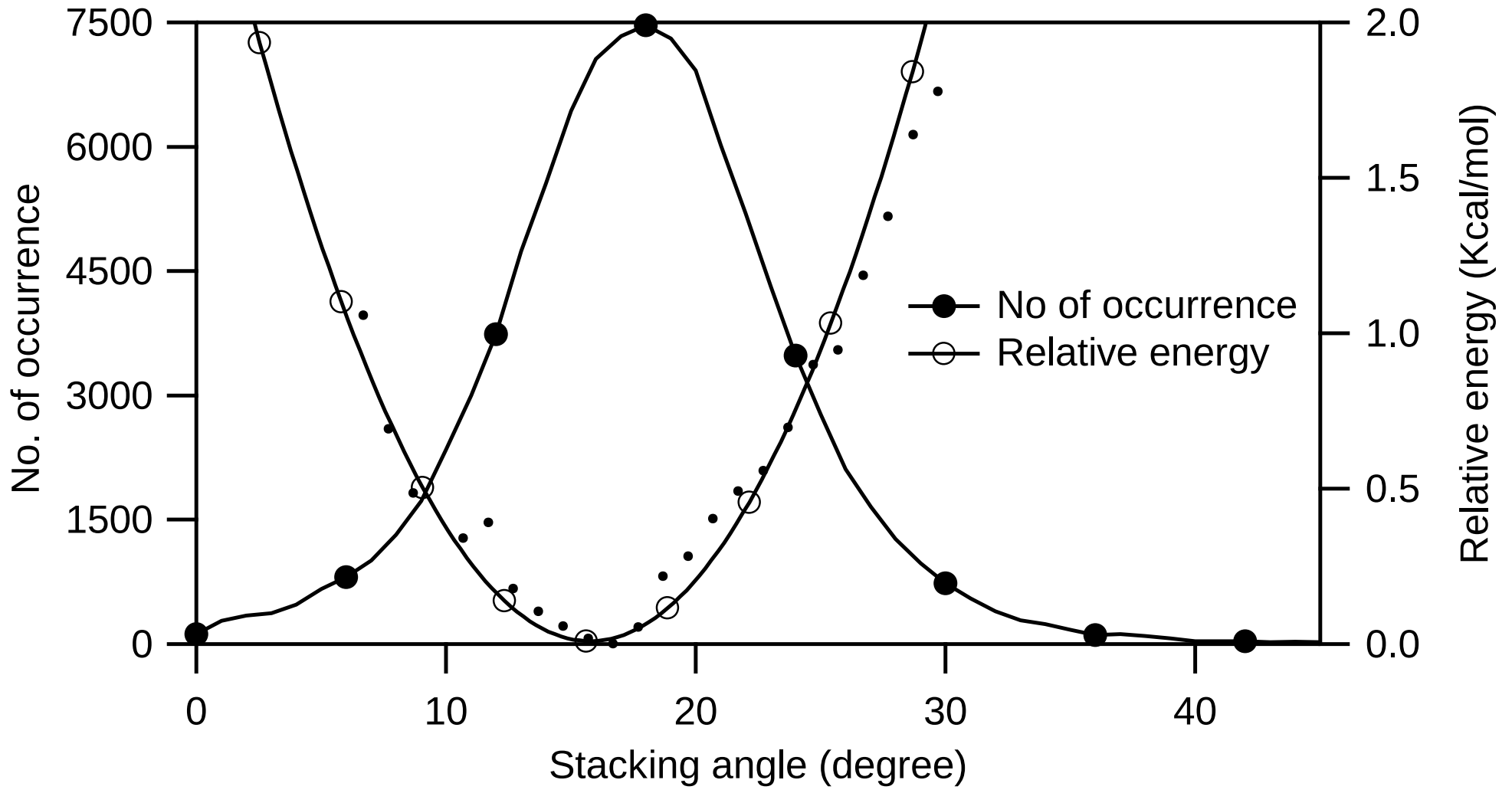

**Figure S4. (b)** The plot between the number of occurrence, potential energy with respect to stacking angle calculation from molecular dynamics simulation for DCV5T-Me molecule

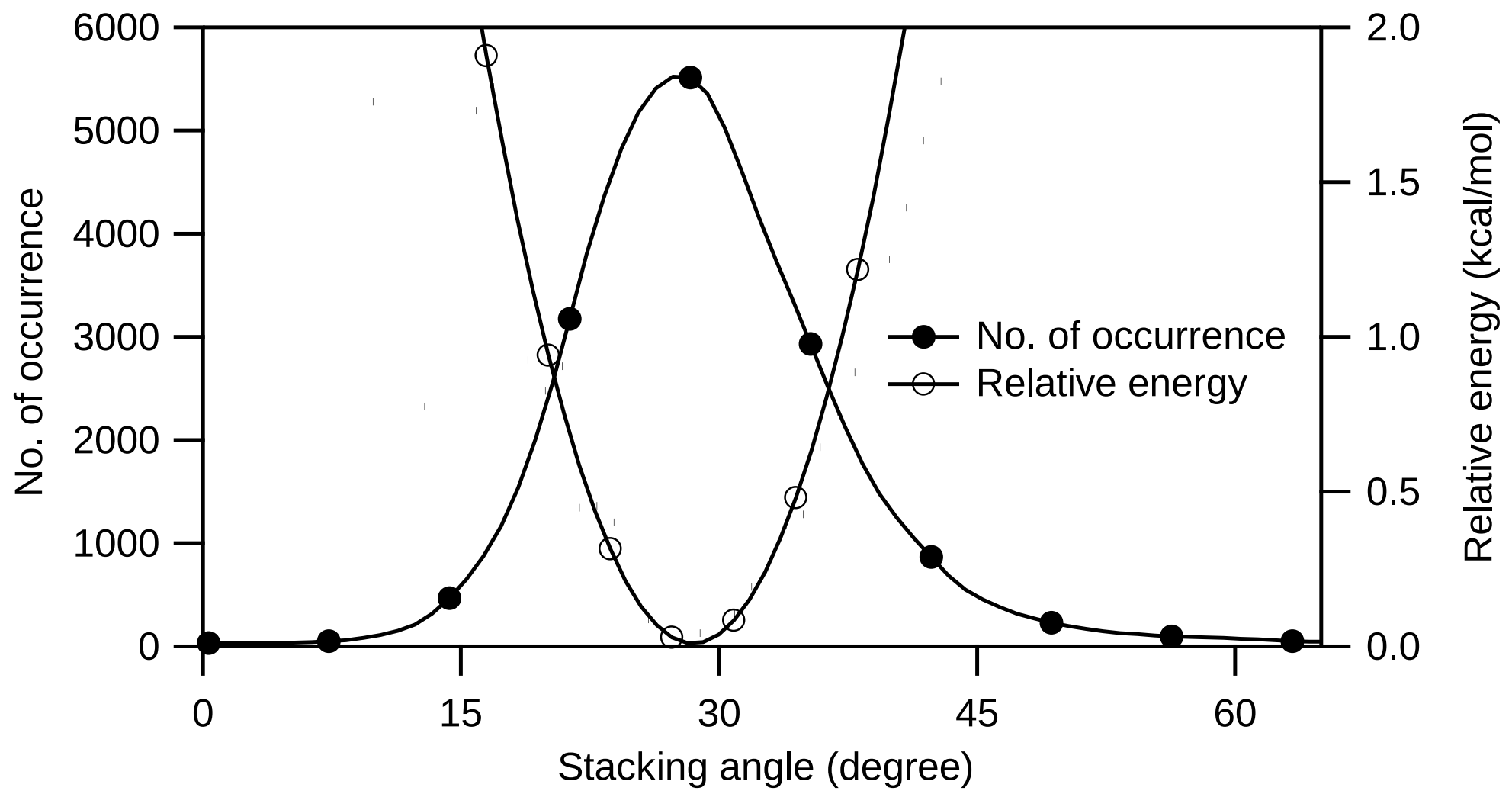

**Figure S5. (a)** The mean squared displacement of positive charge in BDHTT-BBT molecule with respect to time

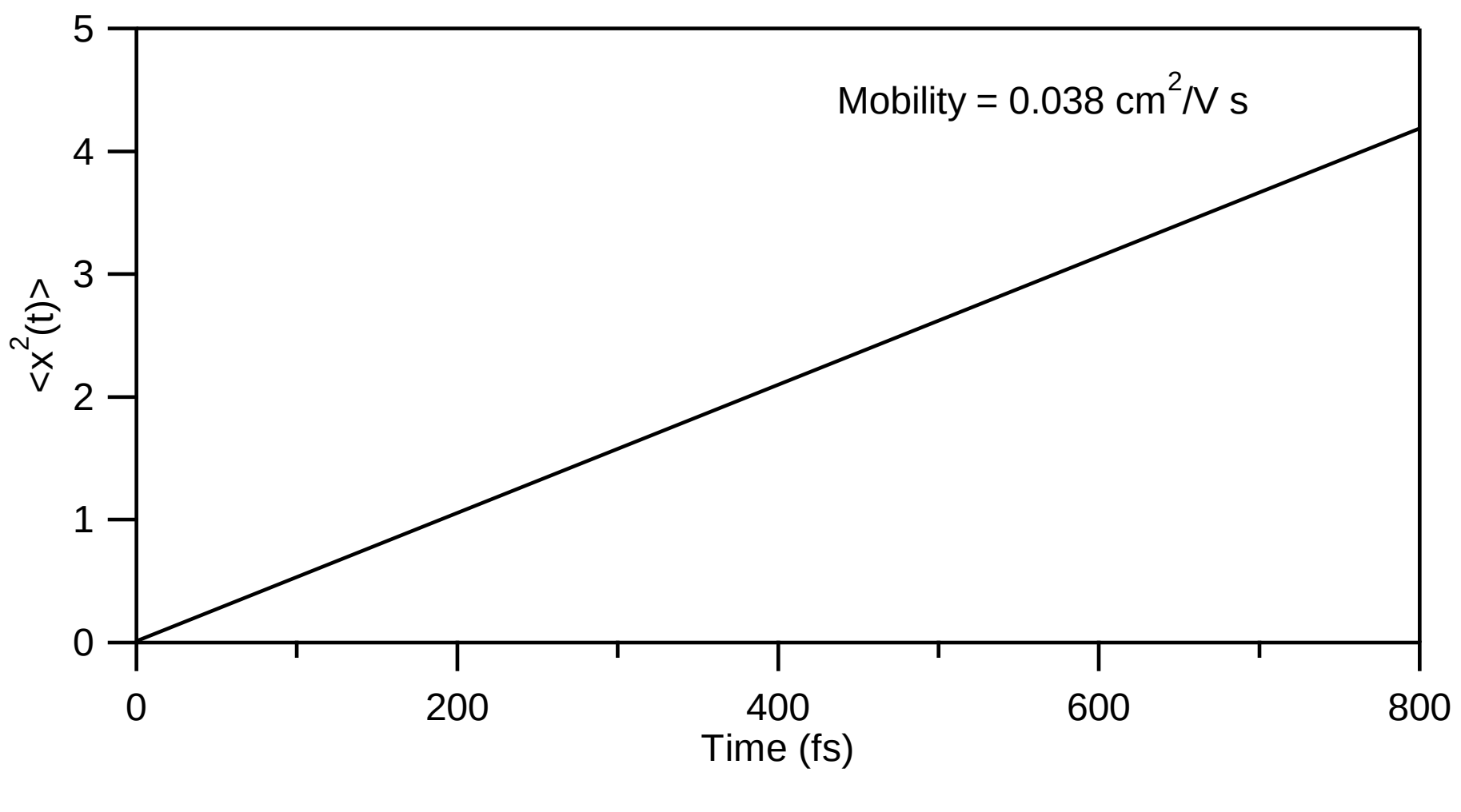

**Figure S5. (b)** The mean squared displacement of positive charge in DCV5T-Me molecule with respect to time

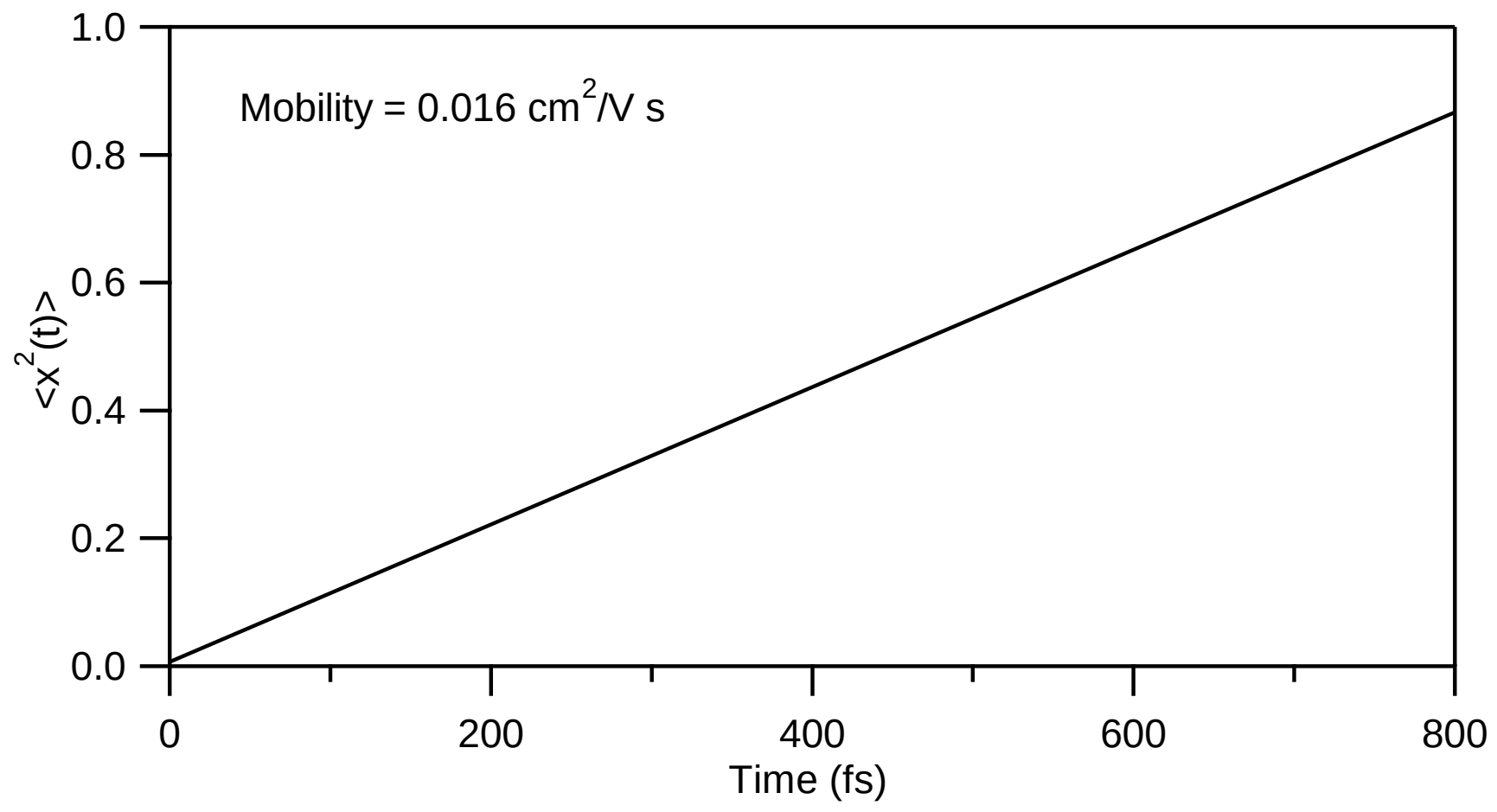

**Figure S5. (c)** The mean squared displacement of negative charge in DCV5T-Me molecule with respect to time

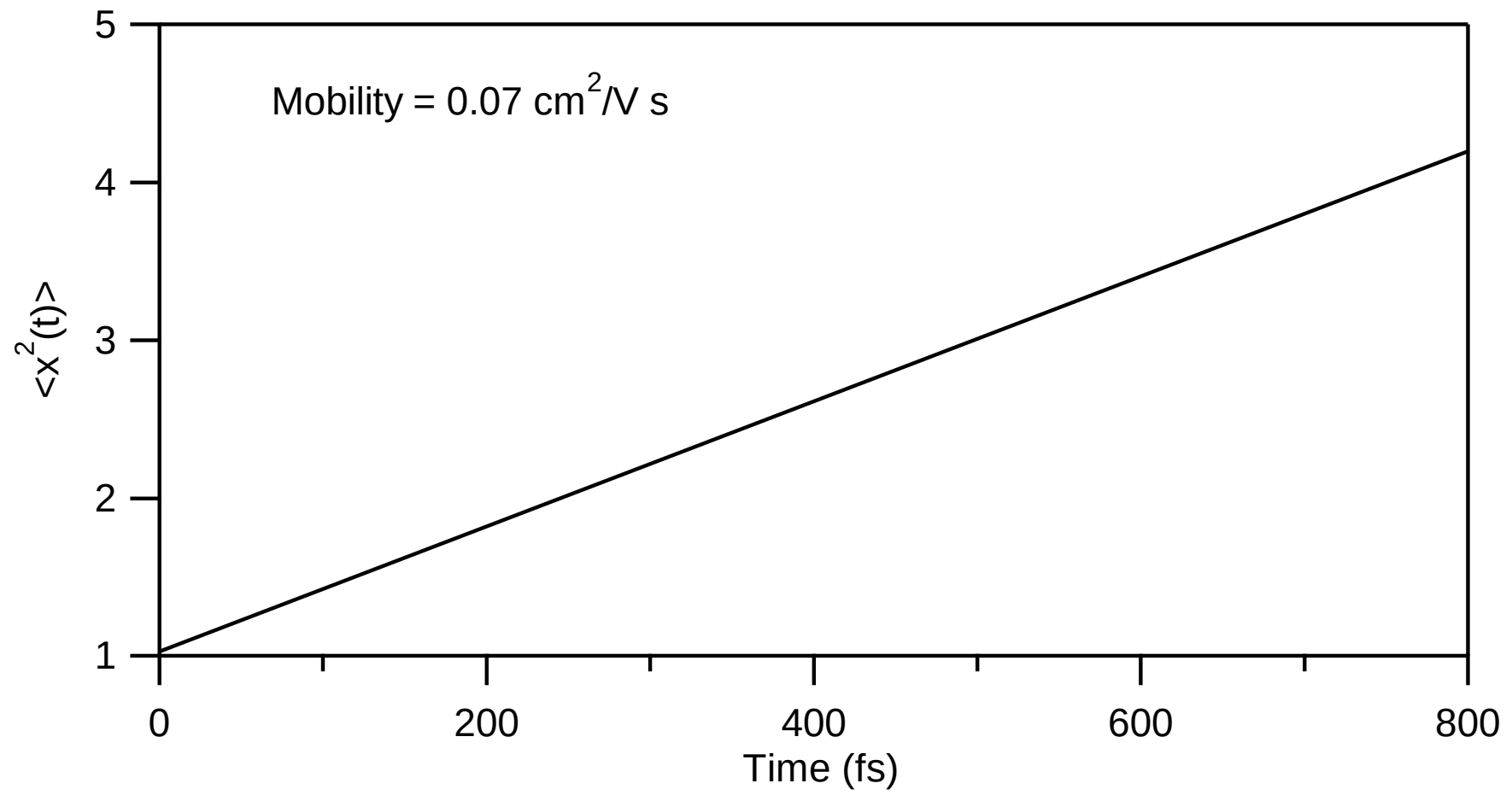

**Figure S6. (a).** The survival probability and disorder drift of positive charge in BDHTT-BBT molecule with respect to time.

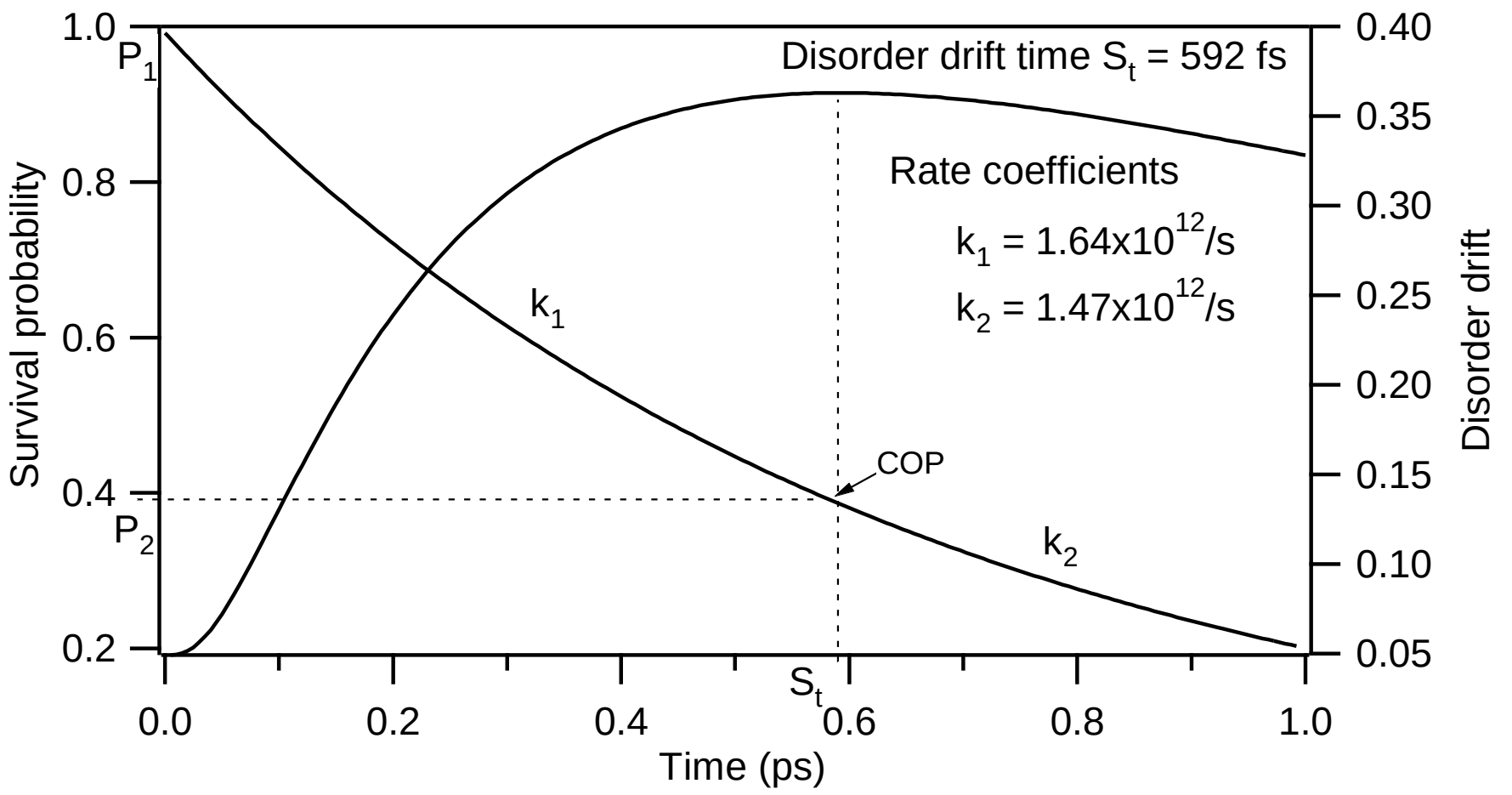

**Figure S6. (b).** The survival probability and disorder drift of negative charge in DCV5T-Me molecule with respect to time.

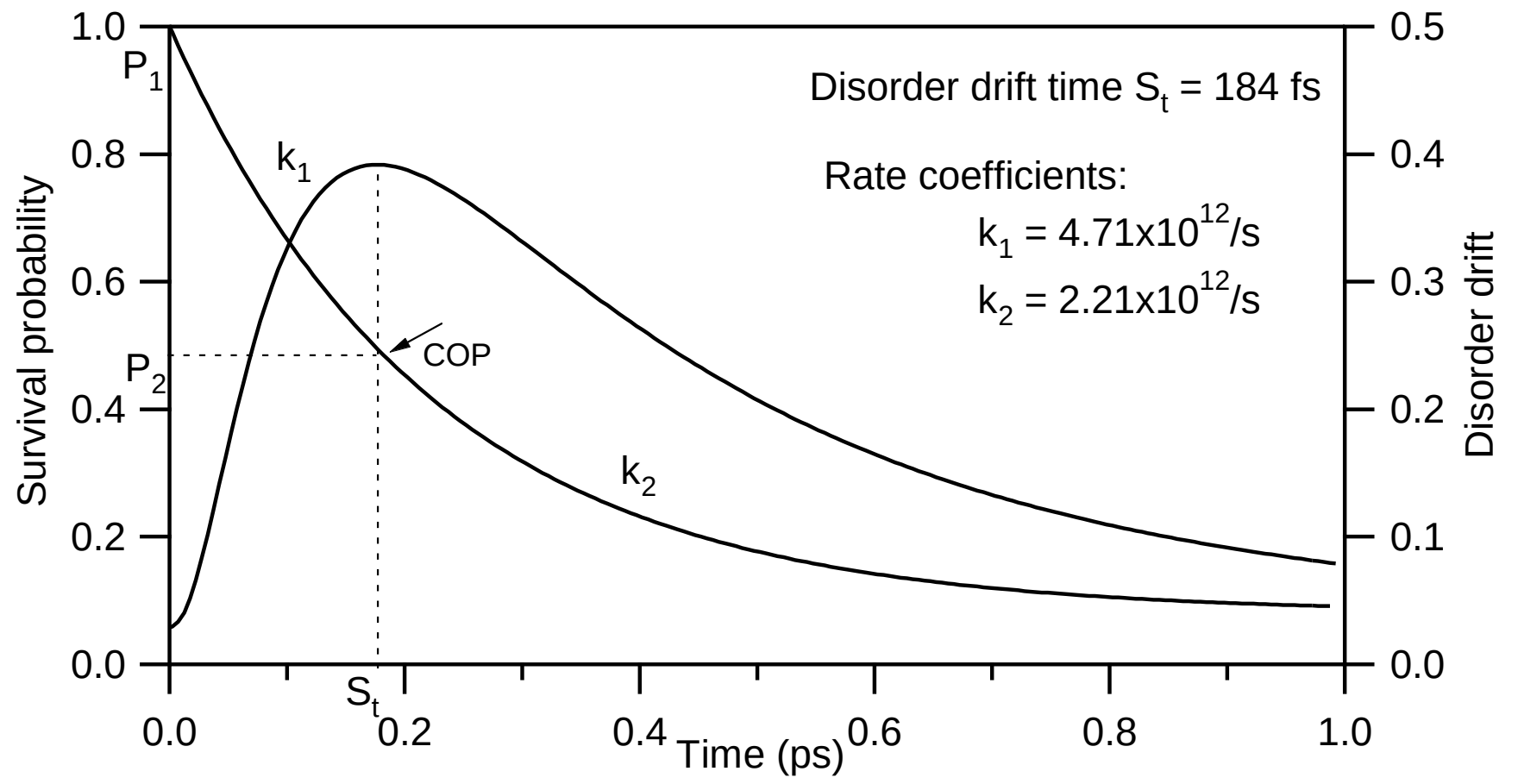

**Figure S7 (a).** Time evolution of the rate coefficient for hole transport in BDHTT-BBT molecule in (a) coherent regime (b) non-coherent regime.

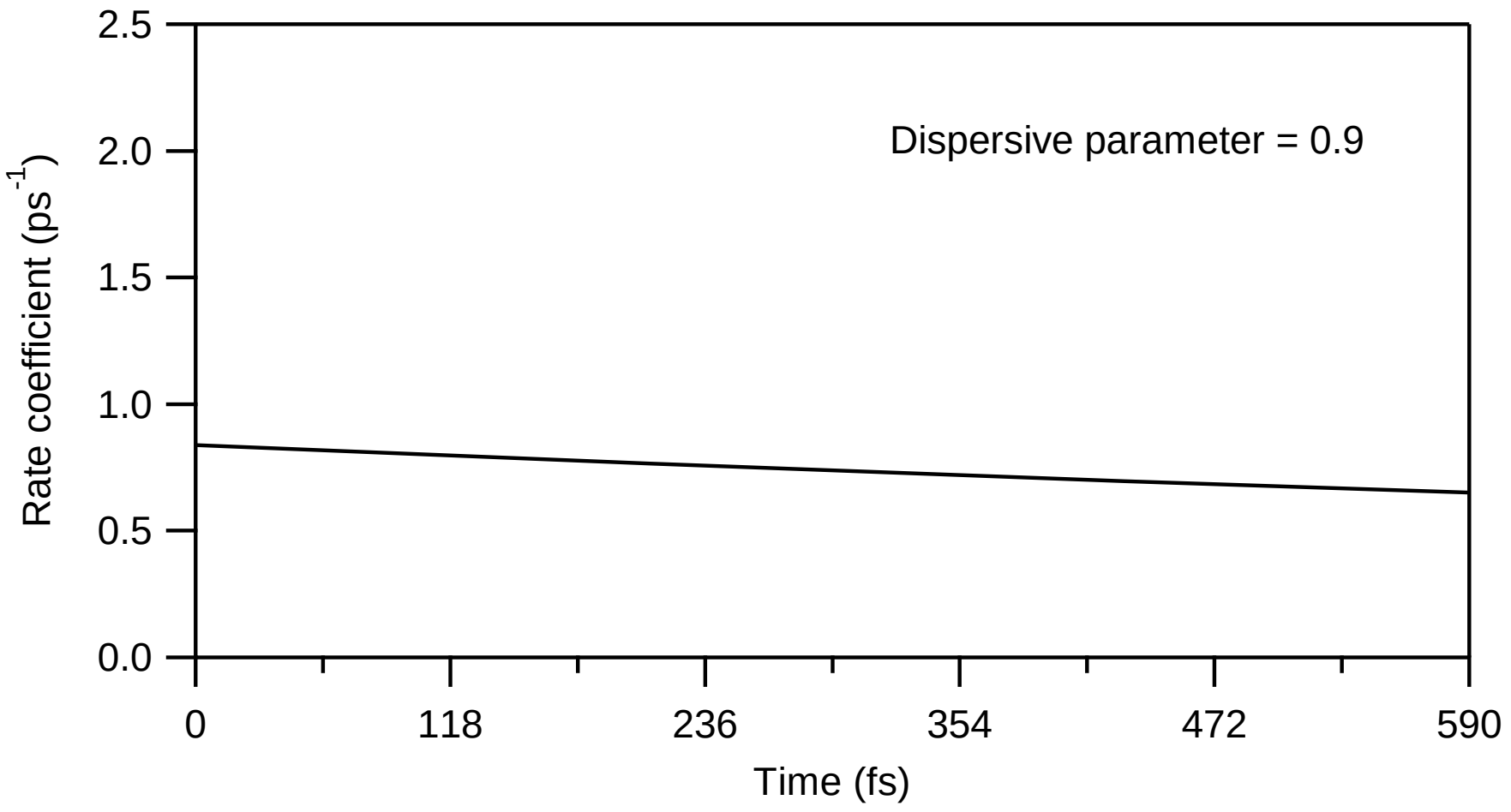

**(a)**

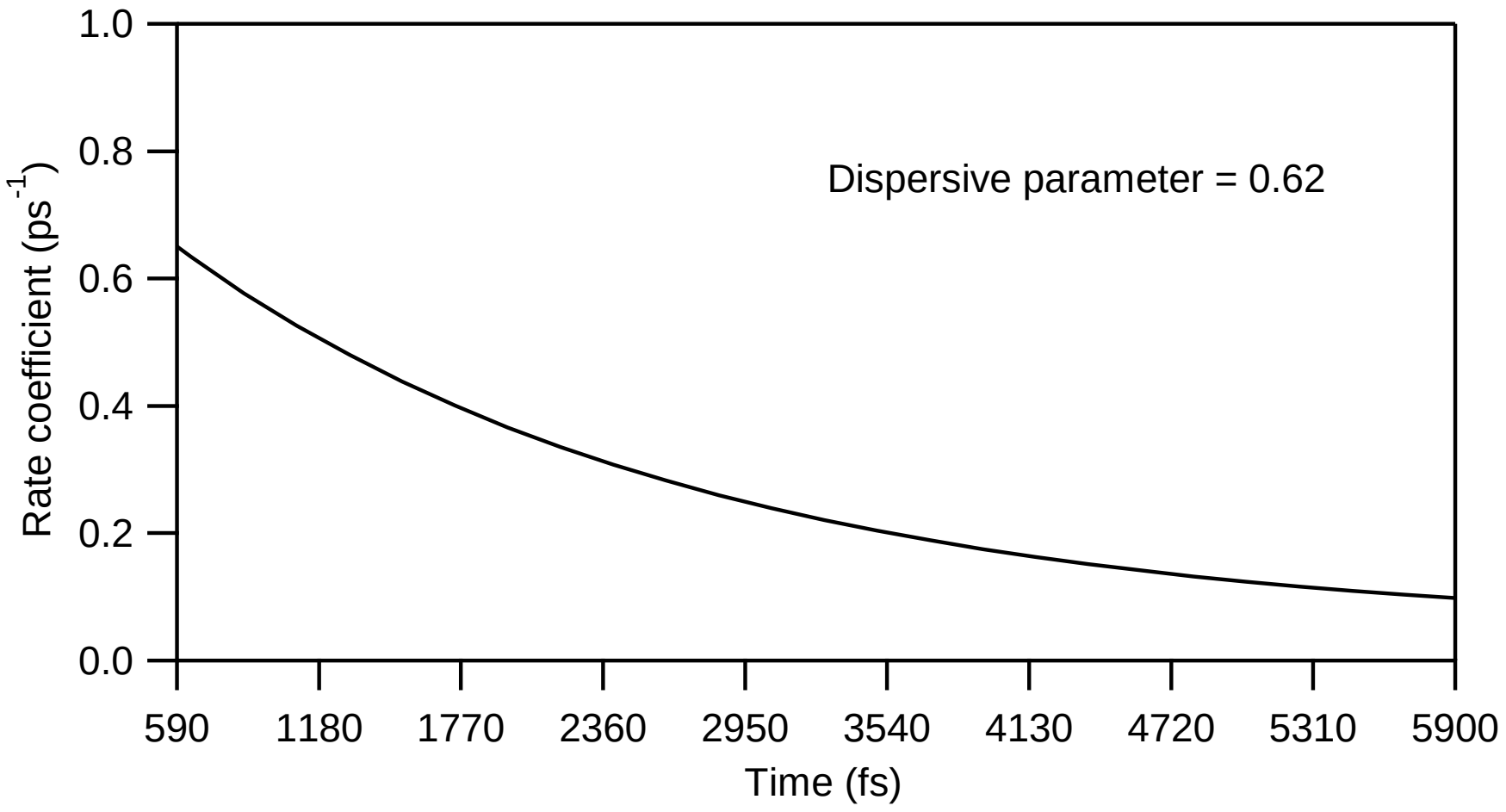

**(b)**

**Figure S7 (b).** Time evolution of the rate coefficient for electron transport in DCV5T-Me molecule in (a) coherent regime (b) non-coherent regime.

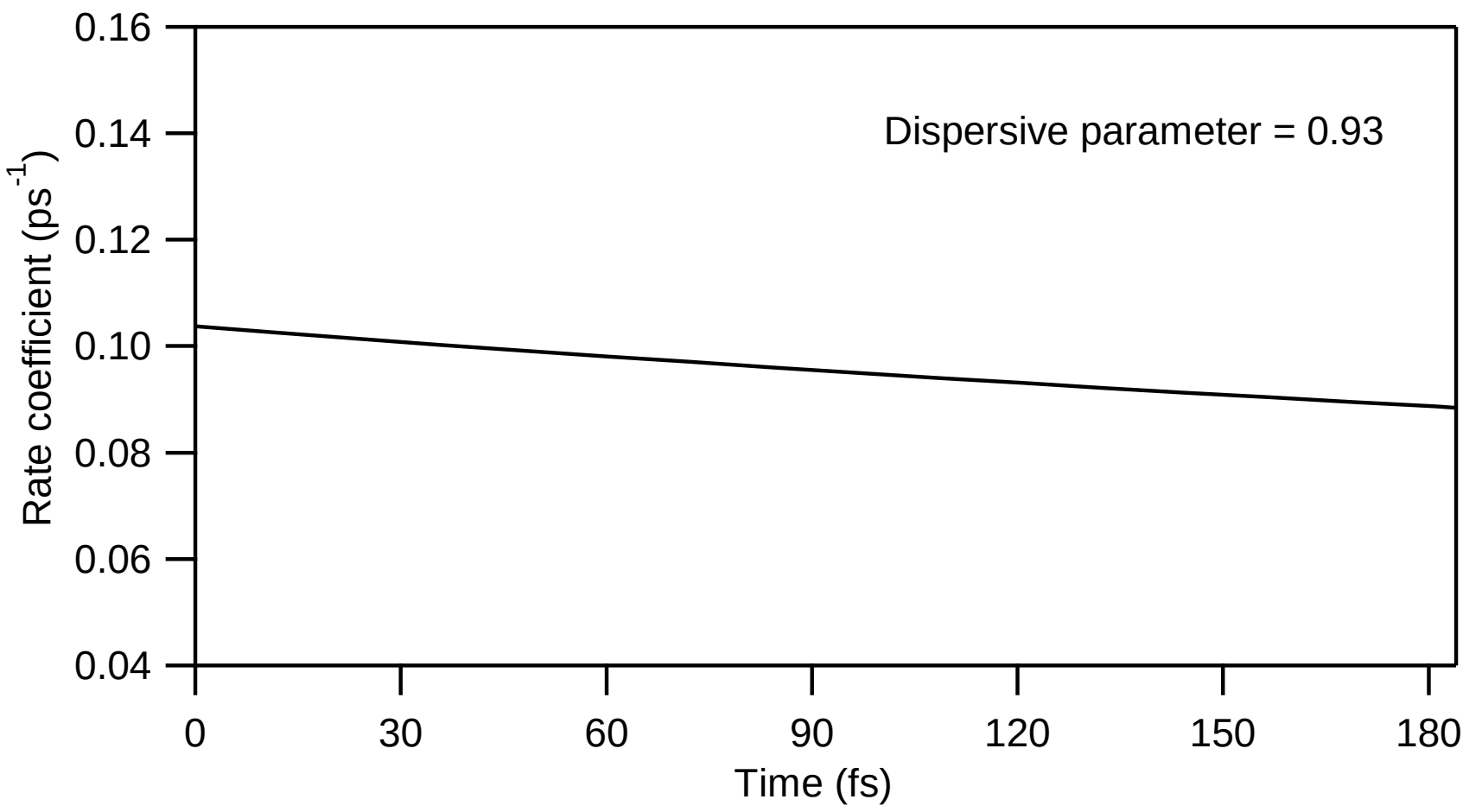

**(a)**

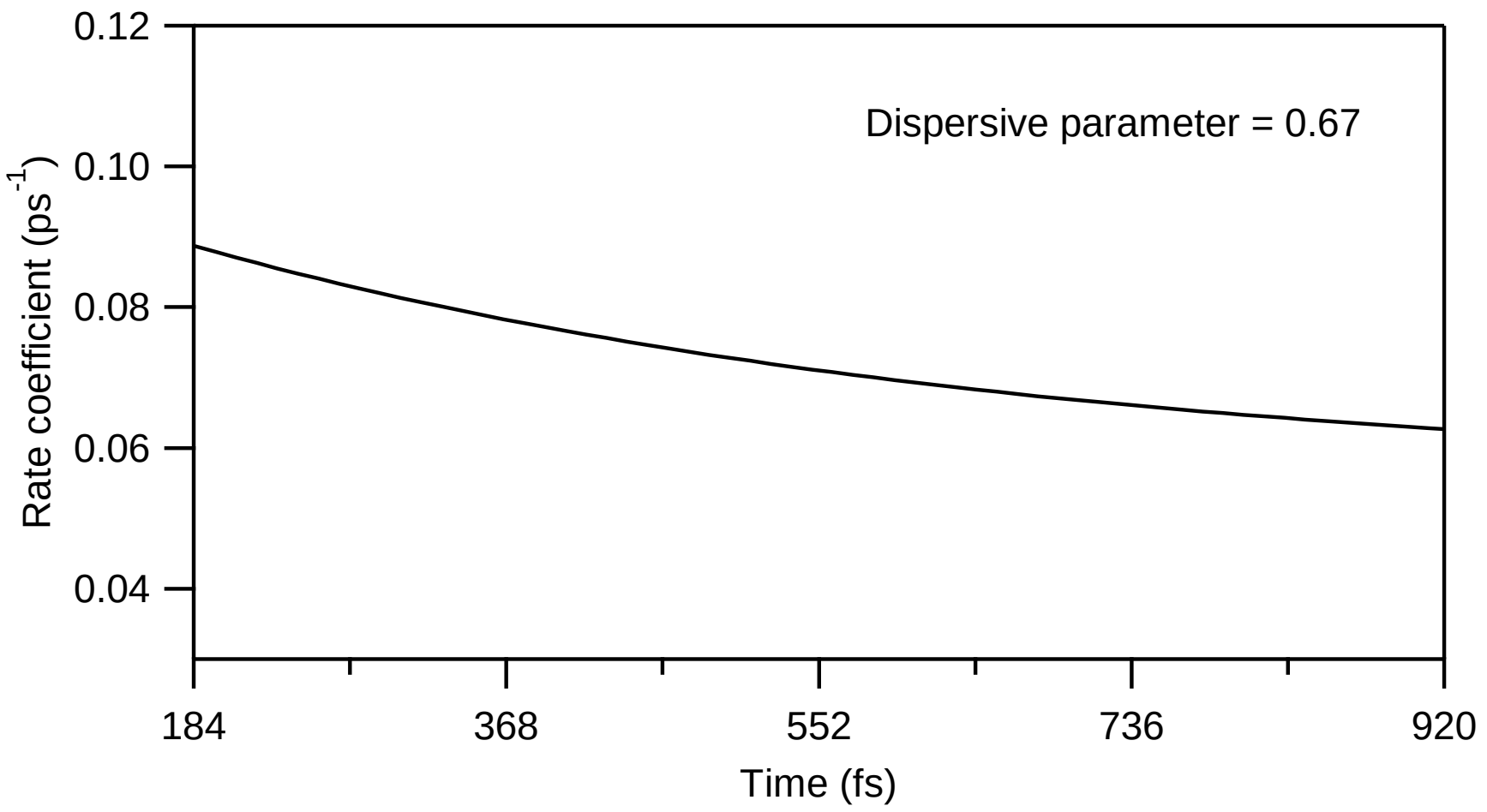

**(b)**

**Figure S8.(a)** Time evolution of momentum ratio and potential distribution for hole transport in BDHTT-BBT molecule

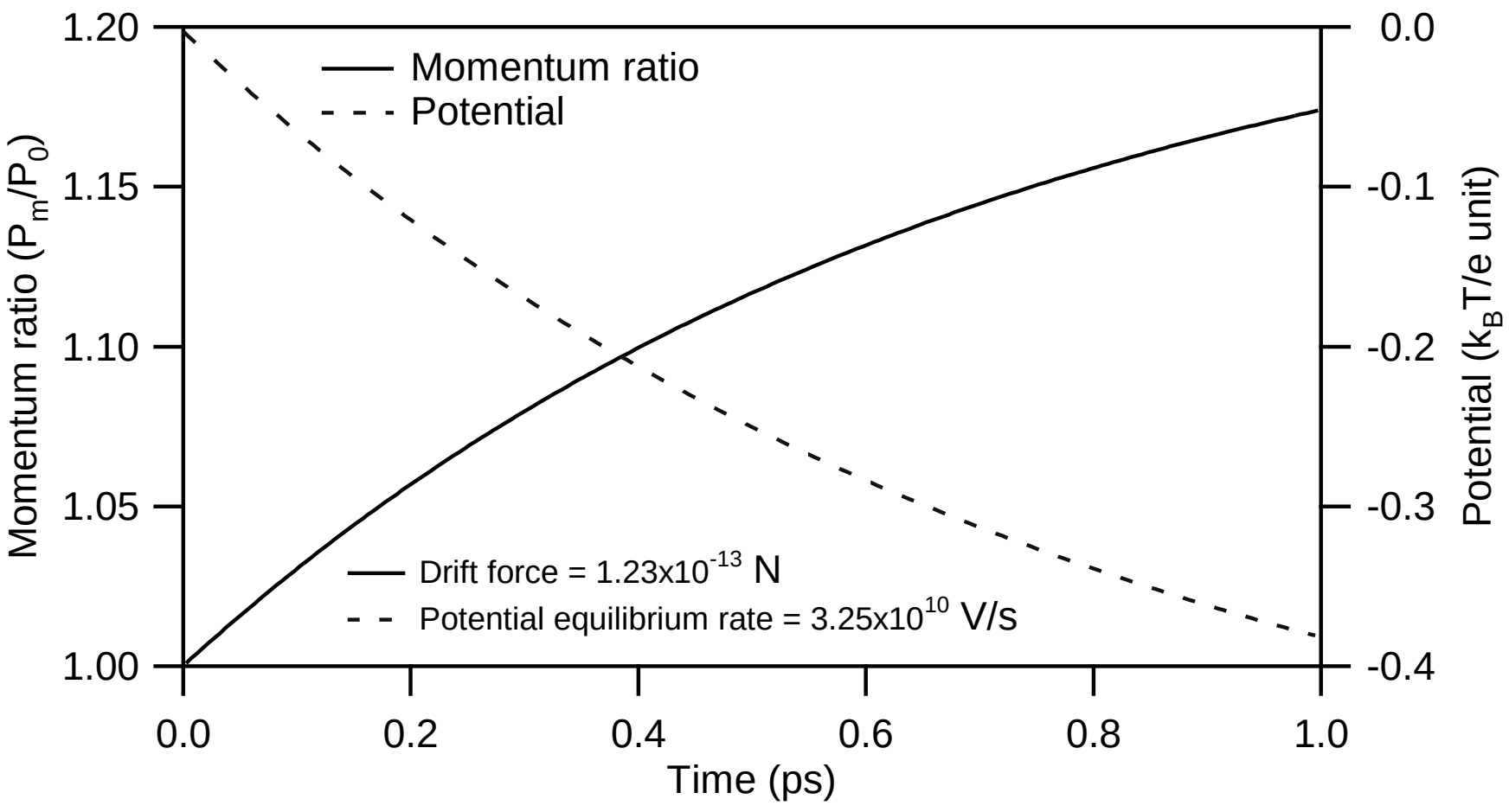

**(a)**

**Figure S8.(b)** Time evolution of momentum ratio and potential distribution for electron transport in DCV5T-Me molecule

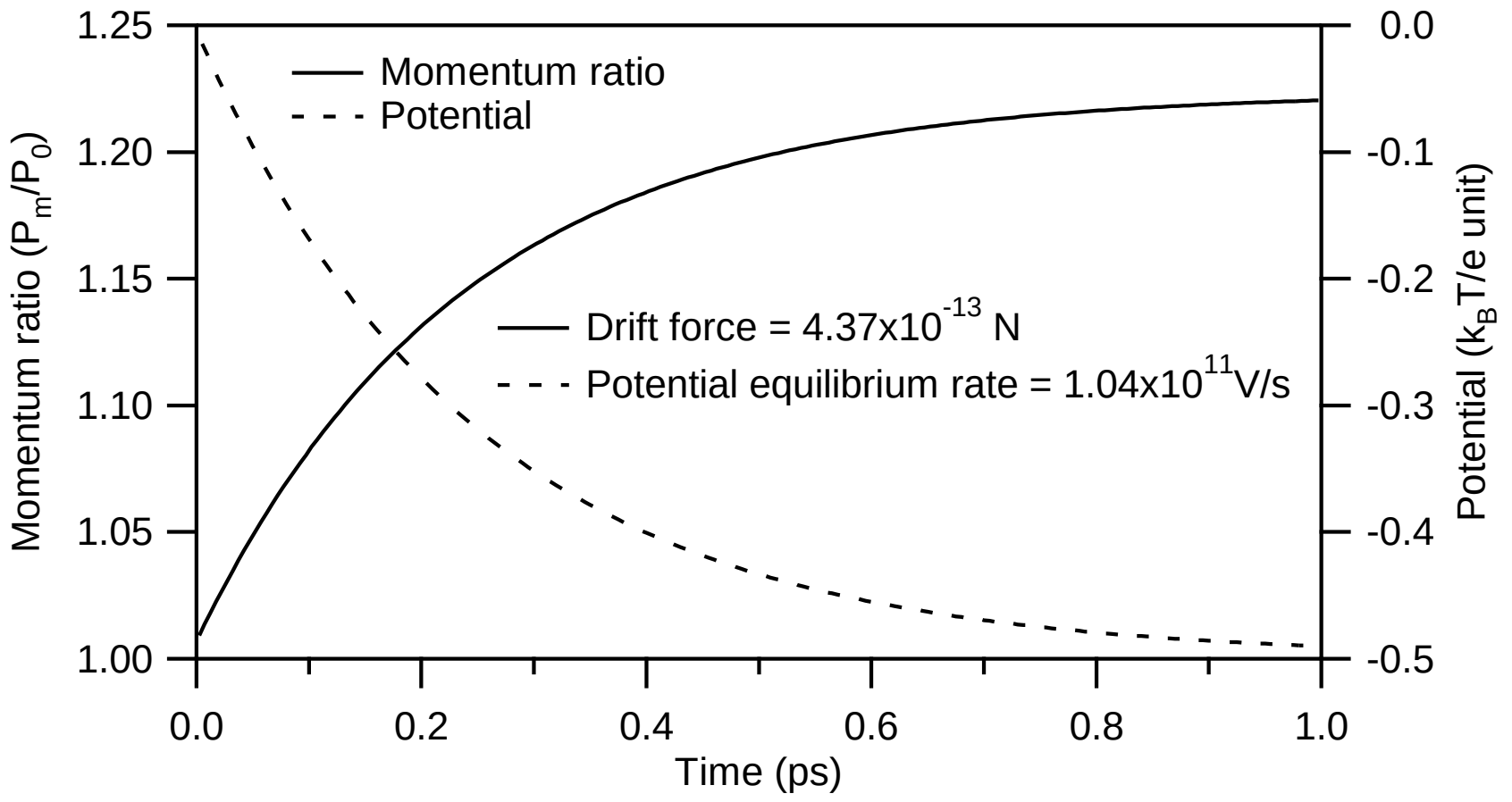

**Figure S9. (a)** Time evolution of the density flux for hole transport in BDHTT-BBT molecule

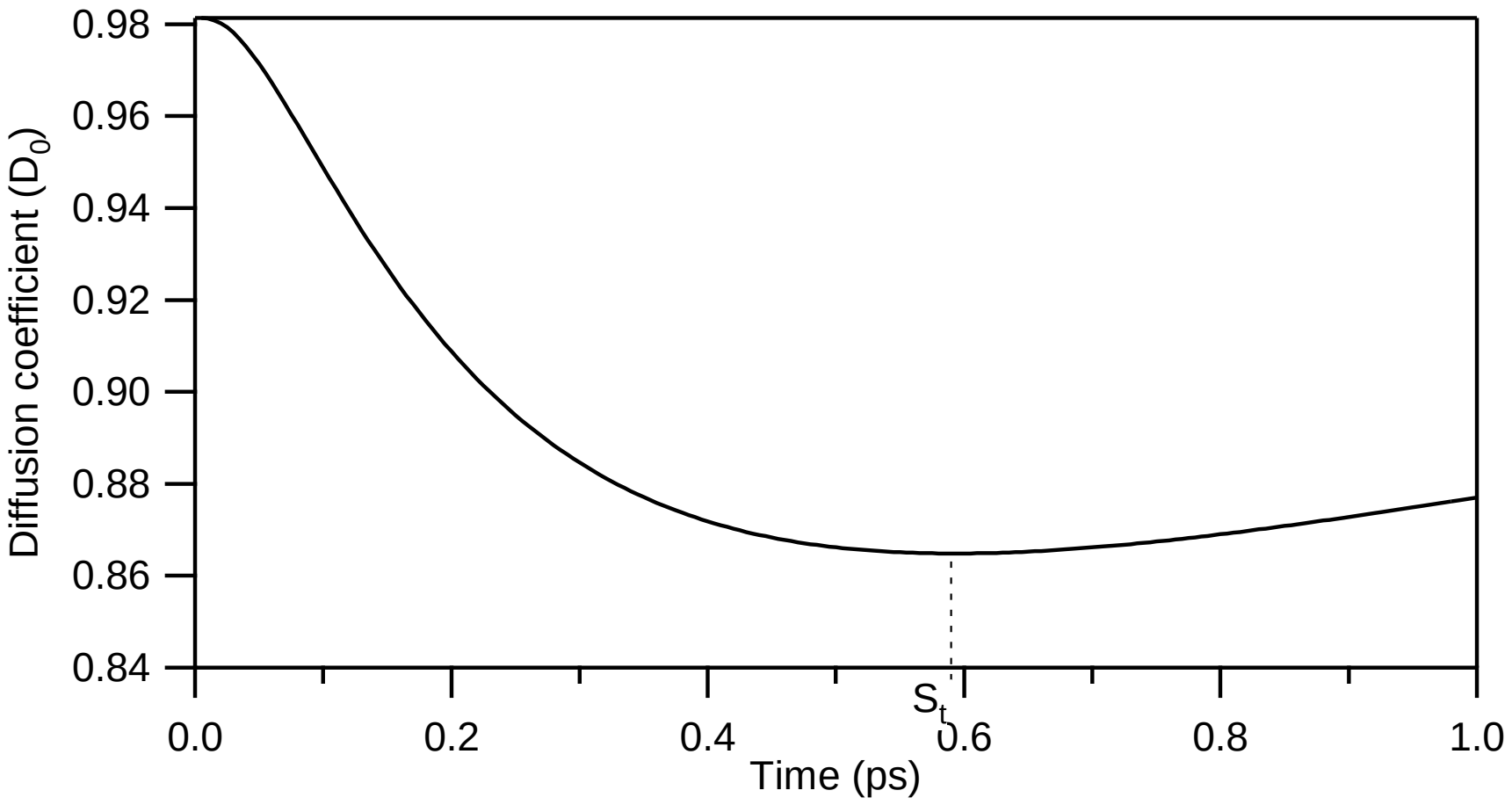

**Figure S9. (b)** Time evolution of the density flux for electron transport in DCV5T-Me molecule

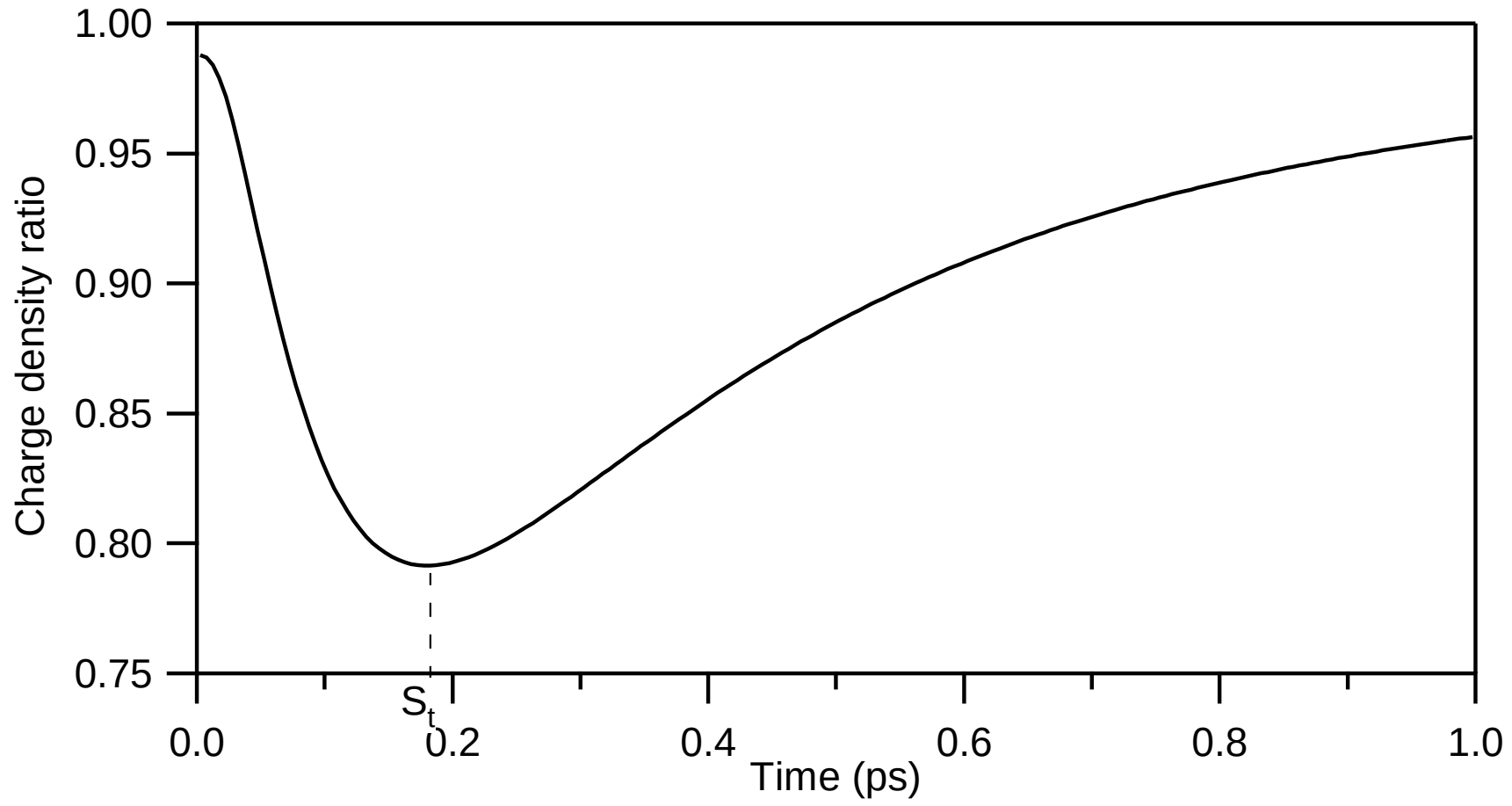

**Figure S10. (a)** Time evolution of the diffusion coefficient for hole transport in BDHTT-BBT molecule

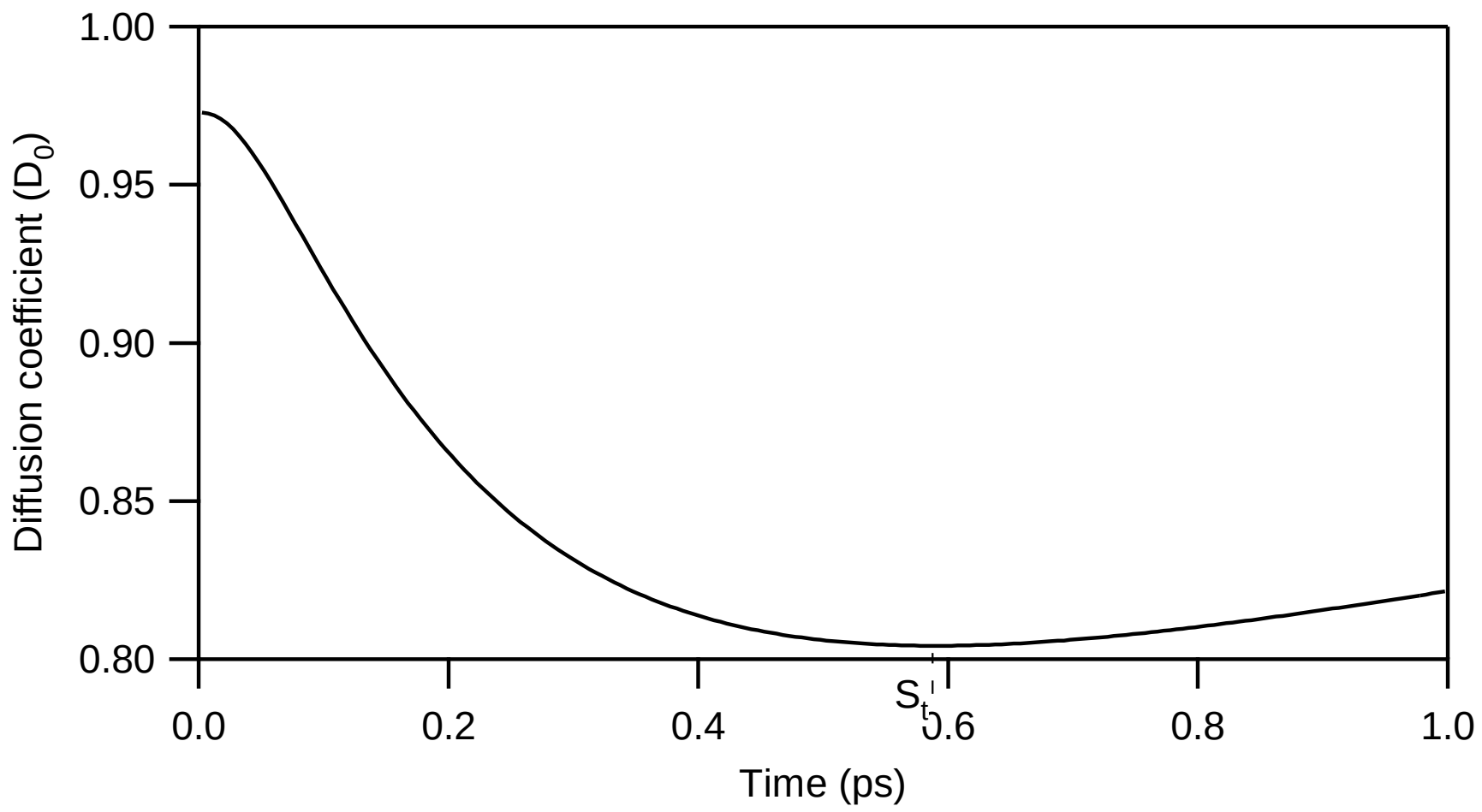

**Figure S10. (b)** Time evolution of the diffusion coefficient for electron transport in DCV5T-Me molecule

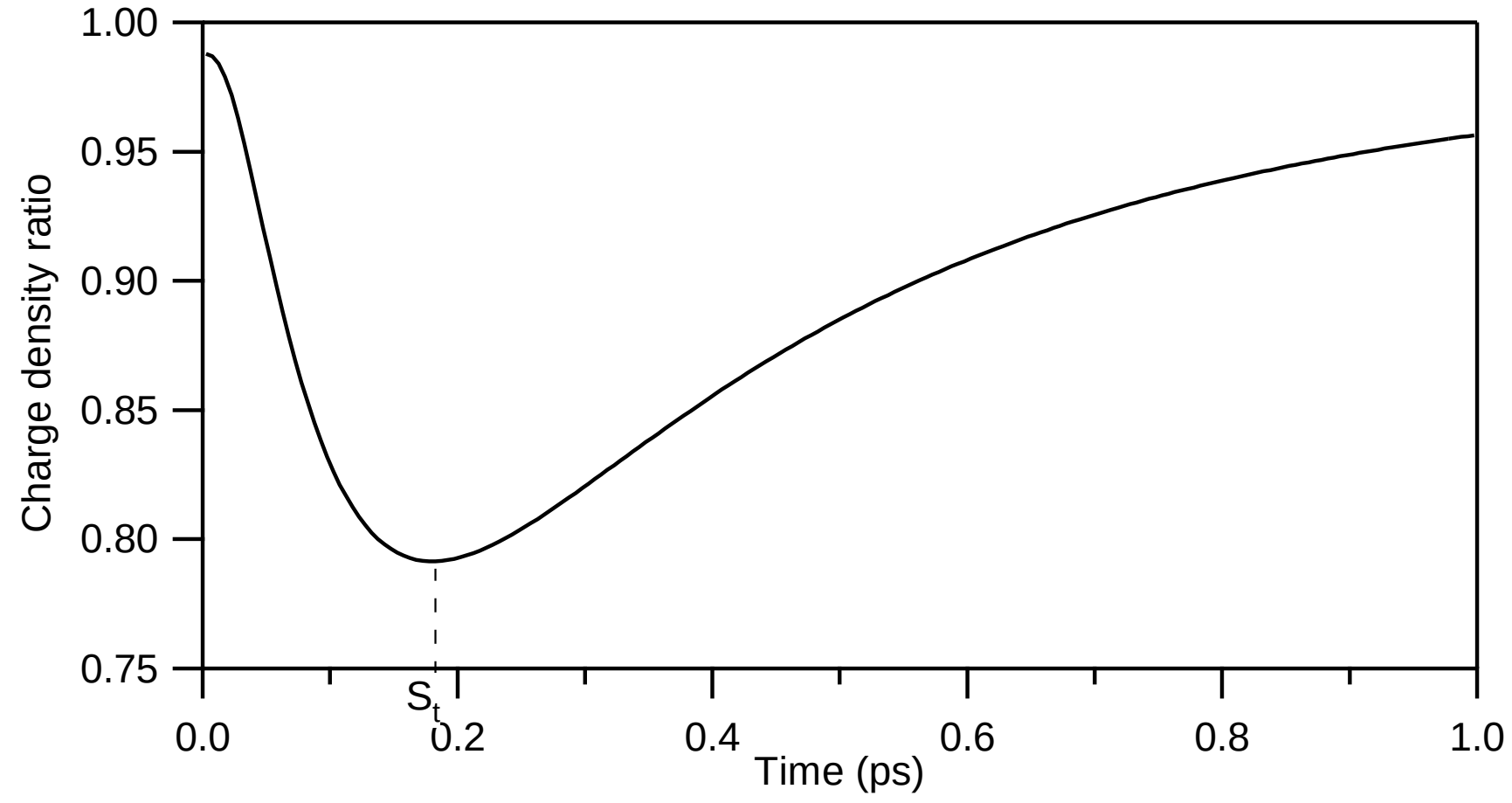

The charge transport properties of BDHTT-BBT and DCV5T-Me molecules are studied by using the charge carrier dynamical parameters, such as charge carrier momentum distribution, rate of change of density flux, drift force, rate of change of potential difference at particular site and diffusion coefficient and are described as follows,

The momentum of the charge carrier ($P_{mom}$) is associated with the charge density $(\rho)$ and is described as, (see *Introduction to Quantum mechanics*, Griffiths, D. J.; *Pearson Education* **2005**, 230-260).

$$P_{mom} = \hbar \left( \frac{3\pi^2 \rho}{e} \right)^{1/3} \qquad \text{(S1)}$$

Time derivative of momentum is

$$\frac{\partial P_{mom}}{\partial t} = \frac{\hbar}{3} \left( \frac{3\pi^2}{e} \right)^{1/3} \rho^{-2/3} \frac{\partial \rho}{\partial t} \qquad \text{(S2)}$$

From Equation S2, the rate of change of density flux is

$$\frac{\partial \rho}{\partial t} = \frac{3}{\hbar (3\pi^2 e)^{1/3}} \rho^{2/3} F_D \qquad \text{(S3)}$$

where, the drift force, $F_D = \frac{\partial P_{mom}}{\partial t}$.

The continuity equation is

$$\frac{\partial \rho}{\partial t} = -\nabla . J \qquad \text{(S4)}$$

where, $J$ is the current density. Since, $J = \sigma E$

$$\frac{\partial \rho}{\partial t} = -\sigma \nabla . E \qquad \text{(S5)}$$

where, $\sigma$ is the conductivity and $E$ is the electric field.

Using Maxwell equation, $\nabla .E = \frac{\rho}{\varepsilon}$ in to Equation S5 we can write

$$\frac{\partial \rho}{\partial t} = -\sigma \frac{\rho}{\varepsilon} \tag{S6}$$

where, $\varepsilon$ is the electric permittivity of the medium.

Now by substituting the Equation S6 in to Equation S2 we get

$$\frac{\partial P_{mom}(t)}{\partial t} = -\frac{\hbar}{3\varepsilon}\left(\frac{3\pi^2 \rho}{e}\right)^{1/3} \sigma \tag{S7}$$

By comparing the Equations S1 and S7, we can write

$$\frac{\partial P_{mom}(t)}{\partial t} = -\frac{P_{mom}(t)}{3\varepsilon} \sigma \tag{S8}$$

For localized charge transport, the conductivity is described as (see Ref. 30, 31),

$$\sigma = \frac{3}{5} \varepsilon \frac{\partial P(t)}{\partial t} \tag{S9}$$

where, $\frac{\partial P}{\partial t}$ is the rate of transition probability (or charge transfer rate). Now by substituting the Equation S9 in to Equation S8, we can write

$$\frac{\partial P_{mom}(t)}{\partial t} = -\frac{P_{mom}(t)}{5} \frac{\partial P(t)}{\partial t} \tag{S10}$$

Or

$$\frac{\partial P_{mom}(t)}{P_{mom}(t)} = -\frac{1}{5} \partial P(t) \tag{S11}$$

By integrating the above equation on both sides, we get

$$P_{mom}(t) = P_{mom,0}\left(\exp(1 - P(t))\right)^{1/5} \qquad \text{(S12)}$$

where, $P_{mom,0}$ is the initial momentum and $P_{mom}(t)$ is the momentum of the charge distribution at time t which is the function of survival probability $P(t)$.

The kinetic energy of the charge carrier is,

$$E_K = \frac{P_{mom}^2(t)}{2m} \qquad \text{(S13)}$$

By substituting Equation S12 in to Equation S13, we get

$$E_K = E_{K,0}\left(\exp(1 - P(t))\right)^{2/5} \qquad \text{(S14)}$$

where, $E_{K,0}$ is the initial kinetic energy.

The energy conservation law is

$$U_0 + E_{K,0} = U + E_K \qquad \text{(S15)}$$

where, $U$ and $U_0$ are the final and initial potential energies.

From Equation S15, we can separate the potential and kinetic energy parts as

$$U - U_0 = E_{K,0} - E_K \qquad \text{(S16)}$$

By substituting Equation S14 in to Equation S16, we can write

$$\Delta U = E_{K,0}\left(1 - \left(\exp(1 - P(t))\right)^{2/5}\right) \qquad \text{(S17)}$$

Note that here the charge transfer is due to thermal diffusion process, and no external electric field is applied. Here, the initial kinetic energy is equl to the thermal energy $\left(E_{K,0} = k_B T\right)$. Therefore, Equation S17 becomes

$$\Delta U = k_B T\left(1 - \left(\exp(1 - P(t))\right)^{2/5}\right) \qquad \text{(S18)}$$

The presence of excess charge on the one end of π-stacked molecular chain introduces the potential difference, $V_d = \frac{\Delta U}{e}$. The charge diffusion will occur up to a point where the potential equilibrium is reached, that is, $V_d$ = 0. During the CT, the change in potential difference with respect to time is defined as,

$$V_d(t) = \frac{k_B T}{e}\left(1 - \left(\exp(1 - P(t))\right)^{2/5}\right) \qquad \text{(S19)}$$

The presence of excess charge on one end of the π-stacked molecular system is responsible for the existence of potential difference and is described through Poisson's equation

$$\frac{\partial^2 V_d}{\partial X^2} + \frac{\rho}{\varepsilon_0} = 0 \qquad \text{(S20)}$$

In this case, charge density is purely depending on dynamic disorder (see Ref. 30) and potential difference is depending on conjugation length (or π-stacking distance). Therefore, the solution of the Poisson equation is

$$V_d = \frac{\rho}{2\varepsilon_0}\left\langle X^2 \right\rangle \qquad \text{(S21)}$$

For one dimensional motion, the mean squared displacement is,

$$\left\langle X^2 \right\rangle = 2Dt \qquad \text{(S22)}$$

Substituting the Equation S22 in to Equation S21, we get

$$V_d = \frac{\rho}{\varepsilon_0} Dt \qquad \text{(S23)}$$

where, $D$ is the diffusion coefficient. By differentiating the Equation S23 with respect to time,

$$\left|\frac{\partial V_d}{\partial t}\right| = \frac{\rho}{\varepsilon_0} D \qquad \text{(S24)}$$

or

$$\frac{\partial V_d}{\partial t} = D\frac{\partial^2 V_d}{\partial X^2} \tag{S25}$$

The above Equation S25 is similar to the diffusion equation. Now by multiplying an electronic charge *e* on both sides of the Equation S25, we get $(E = eV_d)$

$$\frac{\partial E}{\partial t} = \frac{eD}{\varepsilon_0}\rho \tag{S26}$$

To get the expression for $\frac{\partial E}{\partial t}$ we have to use the following thermodynamical equations for energy, pressure and volume, as

$$\partial E = -P_d\, \partial V \tag{S27}$$

where, $P_d$ is the pressure and $\Delta V$ is the change in volume. Now differentiate Equation S27 with respect to time, we get the change in energy with respect to time is (or energy rate)

$$\frac{\partial E}{\partial t} = -P_d \frac{\partial V}{\partial t} \tag{S28}$$

We assume that the change in volume with respect to time is constant, due to the thermal averaging, and rewrite the Equation S28 following way,

$$\frac{1}{P_d}\frac{\partial E}{\partial t} = \frac{\partial\left(\frac{\partial E}{\partial t}\right)}{\partial P_d} = -\frac{\partial V}{\partial t} \tag{S29}$$

By multipling by $\frac{T}{\partial T}$ on both sides of Equation S29, we get

$$\frac{T}{\partial T}\frac{\partial\left(\frac{\partial E}{\partial t}\right)}{\partial P_d} = -\frac{\partial V}{\partial t}\frac{T}{\partial T} \tag{S30}$$

where, T is the temperature. By rearranging the above equation S30, and substitute the thermodynamical relations $\frac{\partial Q}{\partial V} = T\frac{\partial P_d}{\partial T}$ and $\frac{\partial Q}{T} = \partial S$ in to Equation S30, we get

$$\partial\left(\frac{\partial E}{\partial t}\right) = -\partial S\frac{\partial T}{\partial t}, \qquad \text{(S31)}$$

where, the thermodynamic quantities $Q$ and $S$ are referred as quantity of heat energy and entropy. The time evolution on thermal energy, $E = k_B T$ of the system (due to the external interaction and the thermal fluctuation) is described by

$$\frac{\partial E}{\partial t} = k_B\frac{\partial T}{\partial t} \qquad \text{(S32)}$$

By comparing the Equations S31 and S32, we can write

$$\frac{\partial\left(\frac{\partial E}{\partial t}\right)}{\frac{\partial E}{\partial t}} = -\frac{1}{k_B}\partial S \qquad \text{(S33)}$$

By integrating on both sides of Equation S33, we get

$$\left(\frac{\partial E}{\partial t}\right)_S = \left(\frac{\partial E}{\partial t}\right)_{S0}\exp\left(-\frac{S(t)}{k_B}\right) \qquad \text{(S34)}$$

where, $\left(\frac{\partial E}{\partial t}\right)_S$ is the dynamic disorder dependent energy rate, $\left(\frac{\partial E}{\partial t}\right)_{S0}$ is the energy rate without inclusion of dynamic disorder, $S(t)$ is the time dependent entropy and describes dynamic disorder. This is the Equation, S34 for the energy rate (or rate of perturbation) is limited by dynamic disorder.

As described in previous study (see Reference 30), the density flux equation is

$$\rho_S = \rho_{S_0} \exp\left( -\frac{3S}{5k_B} \right) \tag{S35}$$

By substituting Equations S34 and S35 in to Equation S26, we get the expression for diffusion as,

$$D_S = D_{S_0} \exp\left( -\frac{2S}{5k_B} \right) \tag{S36}$$

where, $D_{S_0}$ is the diffusion coefficient for absence of dynamic disorder and $D_S$ is the dynamic disorder dependent diffusion coefficient.